\documentclass[aps,prc,twocolumn,floatfix,showpacs,preprintnumbers,amsmath,amssymb,nofootinbib,groupedaddress]{revtex4-1}
\usepackage{color}
\usepackage{graphicx}
\usepackage{dcolumn}    
\usepackage{multirow}
\usepackage{bm}         
\usepackage{sidecap}
\usepackage{hyperref}   
\usepackage{amsmath}
\usepackage{graphicx}
\usepackage{float}
\usepackage{adjustbox}
\usepackage{caption}
\usepackage{subcaption}

\usepackage{mathtools} 
\usepackage{mhchem} 

\usepackage[dvipsnames,usenames]{xcolor}
\usepackage{mathrsfs,natbib}
\usepackage{epsf,amssymb,amsbsy,amsfonts,amssymb,amsmath}
\usepackage{slashed}
\usepackage{comment}

\usepackage{hyperref}
\usepackage[justification=raggedright]{caption}
\definecolor{dark-red}{rgb}{0.,0.,0}
\definecolor{dark-blue}{rgb}{0.,0.,1}
\definecolor{medium-blue}{rgb}{0,0,1}
\hypersetup{
    colorlinks, linkcolor={dark-red},
    citecolor={dark-blue}, urlcolor={medium-blue}
}

\begin{document}
%
\title{Multi-messenger and multi-physics Bayesian inference for GW170817 binary neutron star merger}

\author{H. G\"{u}ven}
\affiliation{Universit\'e Paris-Saclay, CNRS/IN2P3, IJCLab, 91405 Orsay, France}
\affiliation{Physics Department, Yildiz Technical University, 34220 Esenler, Istanbul, Turkey}

\author{K. Bozkurt}
\affiliation{Physics Department, Yildiz Technical University, 34220 Esenler, Istanbul, Turkey}
\affiliation{Universit\'e Paris-Saclay, CNRS/IN2P3, IJCLab, 91405 Orsay, France}

\author{E. Khan}
\affiliation{Universit\'e Paris-Saclay, CNRS/IN2P3, IJCLab, 91405 Orsay, France}

\author{J. Margueron}
\affiliation{Univ Lyon, Univ Claude Bernard Lyon 1, CNRS/IN2P3, IP2I Lyon, UMR 5822, F-69622, Villeurbanne, France}

\date{\today}

%
\begin{abstract}
The tidal deformability probability distribution extracted from GW170817 alone, or including multi-messenger information, is confronted to astrophysical and nuclear physics additional constraints within a semi-agnostic approach for the dense matter equation of state.
We use Bayesian statistics to combine together low density nuclear physics data, such as the ab-initio predictions based on chiral effective field theory interactions or the isoscalar giant monopole resonance, and astrophysical constraints from neutron stars, such as the maximum mass of neutron stars or the probability distribution function of the tidal deformability $\tilde{\Lambda}$ obtained from the GW170817 event.
The so-called posterior probability distribution functions are marginalized over several nuclear empirical parameters ($L_\textrm{sym}$, $K_\textrm{sym}$, $Q_\textrm{sat}$ and $Q_\textrm{sym}$), as well as over observational quantities such as the $1.4M_\odot$ radius $R_{1.4}$ and the pressure at twice the saturation density $P(2n_\textrm{sat})$. The correlations between $L_\textrm{sym}$ and $K_\textrm{sym}$ and between $K_\textrm{sat}$ and $Q_\textrm{sat}$ are also further analyzed.
Tension is found between the posteriors: the first one is localized in the tidal deformability probability distribution itself, depending whether multi-messenger analysis is included or not, and the second one is between the observational data and the nuclear physics inputs.
These tensions impact the predictions for $L_\textrm{sym}$, $K_\textrm{sym}$ and $R_{1.4}$ with centroids which differ by 2-3$\sigma$.
Implications for the nuclear equation of state are also discussed.
\end{abstract}




\maketitle

\section{Introduction}\label{int}

While experiments in finite nuclei probe densities around saturation density of nuclear matter
($n_\mathrm{sat}\approx 0.16$~fm$^{-3}$, $\rho_\mathrm{sat}\approx 2.7\times10^{14}$~g/cm$^{3}$)
and heavy-ion collisions explore a wider domain of densities with small isospin asymmetries, neutron stars (NS) are the solely system which explore the equilibrium properties of dense matter at densities well above saturation density and isospin asymmetries close to pure neutron matter~\cite{NewCompStar2019}.
NS physics addresses thus one of the most fundamental question in nuclear physics which is the understanding of the nuclear interaction in dense medium as a function of the density and the isospin asymmetry.
They are excellent systems where the high density behavior of the nuclear equation of state (EoS) can potentially be determined.

However, the difficulty with the analysis of astrophysical observations is that they often carry global information requiring the understanding of many ingredients, such as general relativity, plasma physics, magnetic fields, nuclear and hadron physics, neutrino transport properties, and so on~\cite{NewCompStar2019}.
At variance with astrophysics, experimental conditions in laboratories are usually better controlled.
Therefore, linking theoretical modelings and data requires specific approaches which take into account the specificities of astrophysics and nuclear physics.
We propose an approach combining a semi-agnostic meta-modeling for the nuclear equation of state~\cite{Jerome2018p1} and a Bayesian statistical analysis.
In this way, it is possible to put together data or constraints from very different origins, where the individual impact of the various set of constraints could also be analyzed separately~\cite{Raithel2016,Raithel2017,Jerome2018p2,Lim2018,Lim2019}.

From the nuclear physics side, we consider the many-body perturbation theory (MBPT) predictions in symmetric (SM) and neutron matter (NM) based on $\chi$EFT interactions from the Ref.~\cite{Drischler2016} as a good representation of the present nuclear physics knowledge on the EoS.
These $\chi$EFT interactions include not only two body nucleon-nucleon force but also three body interactions and they reproduce experimental data such as the charge radius, the neutron radius, the weak form factor, and the dipole polarizability of $^{48}$Ca~\cite{Hagen2016}.
The generated EoS will therefore be evaluated with respect to their proximity to the $\chi$EFT band.
We complement this constraint by the experimental information on the Isoscalar Giant Monopole Resonance (ISGMR), which provides a constraint on the parameter $M_c$ defined below saturation density~\cite{Khan2012,Khan2013}.
The nuclear parameter $M_c$ strongly constrains the density dependence of SM below saturation density and happens to be less model-dependent than the usual nuclear empirical parameter (NEP) $K_{sat}$.
The marginalization of the Bayesian probability over the NEP will show the consistency or the tension existing among the various constraints considered here.

From the astrophysical side, the observation of NSs allows to set limits on the maximum mass, for which radio observations only provide a lower bound.
The maximum mass of neutron stars, which impacts the maximum density of stable baryonic matter, fixes the mass boundary between NSs and black holes, which give clues on the understanding of supernova core-collapse mechanism~\cite{Janka2007} as well as of the fate of NS mergers as kilonovae~\cite{Metzger2017}.
The observed masses vary from $1.174(4)M_\odot$~\cite{Ozel2016,Fonseca2016} to about $2M_\odot$~\cite{Antoniadis2013,Ozel2016}.
The well established upper mass limits are: $1.908(16)M_\odot$ for PSR J1614-2230~\cite{Arzoumanian2018}
and $2.01(4)M_\odot$ for PSR J0348+0432~\cite{Antoniadis2013}.
Recently, two new observations have raised up the upper limit to $M_{max}=2.14^{+0.10}_{-0.09}M_\odot$ from Shapiro delay associated to the MSP J0740+6620~\cite{Cromartie2019} and $M_{max}=2.27^{+0.17}_{-0.15}M_\odot$ from magnesium lines associated to the "redback" PSR J2215+5135~\cite{Linares2018}.
To be compatible with observations, we consider that the maximum mass to be reached by the EoS models should lie above the measured centroid mass minus twice the error-bar (95\% confidence level).
In the present work, we fix this limit to be $M_\mathrm{max}^\mathrm{obs}\approx 2M_\odot$.

In 2017, the first gravitational waves (GW) from a binary NS (BNS) merger (GW170817), been detected by the LIGO-Virgo collaboration~\cite{Ligo2017,Ligo2019},
has provided an estimation of the NS tidal deformability $\tilde{\Lambda}$~\cite{Hinderer2008,Hinderer2008v2,Damour2009}.
The tidal deformability is similar to the measure of compactness~\cite{Ligo2017}, and together with a measure of the mass, can be used to extract the NS radius~\cite{Tews2019}.
The tidal deformability extracted from GW170817 is $70<\tilde{\Lambda}<720$ at 90\% confidence level from Ref.~\cite{Ligo2019}, and
$70<\tilde{\Lambda}<500$ from Ref.~\cite{De2018}.
Moreover the $\tilde{\Lambda}$ probability distributions function (PDF) exhibit an interesting structure, doubly peaked from Ref.~\cite{Ligo2019} (with a large and a small peak) and only singly peaked from Ref.~\cite{De2018}.
In the present work, we shall perform a Bayesian analysis exploring the impact of these two different PDF on our results.

The GW170817 signal has been confronted to various nuclear modelings,
going from the most agnostic ones,
such as piece-wise polytropes (PE)~\cite{Raithel2017,Annala2018,Most2018,Fasano2019} and sound speed (CSS) EoS~\cite{Tews2019,Tews2018},
semi-agnostic approaches where matter composition is known, Taylor-Expended (TE) EoS~\cite{Tews2018,Lim2018,Tews2019,Carson2019,Lim2019})
or more traditional approaches based on nuclear interactions or Lagrangians, such as Skyrme-Type Functional (STF)~\cite{Malik2018,Kim2018,Zhou2019,Malik2019,Carson2019}, and Relativistic Mean Field (RMF)~\cite{Hornick2018,Carson2019,Malik2018,Lourenco2019,Nandi2019}.
In Refs.~\cite{Hornick2018,Lourenco2019}, based on RMF modeling the authors
concluded that the NEP $L_{sym}$ is independent of the radius at $1.4M_\odot$ and that most of the explored EoSs are inside the tidal deformability limit ($\tilde{\Lambda}<720$).
In Refs~\cite{Kim2018} and \cite{Malik2019}, 5 and 28 STFs were analyzed predicting
NS radii to be $11.8 \leq R_{1.4}\leq 12.8$~km~\cite{Kim2018} ($R_{1.4}=11.6\pm 1$~km~\cite{Malik2019}) and the tidal deformability for canonic NS mass ($1.4M_\odot$) $308<\Lambda_{1.4}<583$.
Additionally, it is suggested that the ISGMR constrains the NS compactness~\cite{Malik2019}.
Therefore in the present work we investigate the role of the ISGMR to constrain NS EoS.
In Refs.~\cite{Annala2018,Most2018}, PEs were used to calculate NS EoS leading to $12\leq R_{1.4}\leq13.7$~km for the canonical $1.4M_\odot$ NS radius.
Similar results are found using both RMF and STF~\cite{Malik2018,Lim2018,Carson2019} , and TE~\cite{Lim2018,Carson2019}.
Contrary to Ref.~\cite{Hornick2018}, TE EoS from Ref~\cite{Carson2019} showed that
the tidal deformability constrains both the incompressibility slope at the saturation density $M_0$ and $L_{\textrm{sym}}$.
Recently, GW70817 has been reanalysed based on an agnostic approach (CSM) and including a constraint on the maximal mass of NS~\cite{Capano2019}.
This analysis concluded that the NS radius shall be $\sim 11\pm1$~km.
We will see that we come to a similar conclusion in our analysis based on the $\tilde{\Lambda}$ PDF from Refs.~\cite{Ligo2019,De2018}.

In addition to the GW signal, the GW170817 BNS merger have produced an observed electromagnetic (EM) signal (AT2017gfo) and a gamma-ray burst (GRB170817A).
These additional signals are influenced by the properties of the in-spiral NS, and could potentially also help the characterization of the tidal deformability.
A recent multi-messenger Bayesian analysis has been performed based on the present knowledge and modeling of the EM and GRB signals~\cite{Coughlin2019}.
This analysis suggests that $\tilde{\Lambda}\ge 300$~\cite{Coughlin2019}. While one should expect improved modeling of the EM and GRB emission before strong conclusions can be drawn, this analysis illustrates how a global understanding of the transient event could shed light on the estimation of the tidal deformability. However, constraining GW with its counterpart AT2017gfo signal relies on a number of hypotheses which, for some of them, are still under debate. We refer for instance to the recent update on this topic presented in Ref.~\cite{Kiuchi2019}.
In the present work, we confront the suggestion of the $\tilde{\Lambda}$ PDF from Ref.~\cite{Coughlin2019} with the ones based on only the GW signal~\cite{Ligo2019,De2018}.
In this way, we could analyse the sensitivity of the predictions to the $\tilde{\Lambda}$ PDF.

The radii of NS can also be inferred from X-ray emission, whether their are thermal emission from qLMXB (quiescent Low-Mass X-ray Binary) or
X-ray burst~\cite{Steiner2010,Guillot2013,Steiner2014,Ozel2016,Raithel2016,Marino2018,Baillot2019}.
The predictions from these analyses are becoming more and more accurate since the modeling is improving and more and more statistics is being accumulated.
These analyses however require a clear knowledge on some NS properties, such as the composition of their atmosphere, the hydrogen column density on the line site, and in some cases of the magnetic fields ~\cite{Ozel2016}. In the absence of pulsation, the uncertainty on the NS spin could also bring some uncertainties as well.
The traduction of all these uncertainties in the inferred radius is expected to be about 1 to $2$~km~\cite{Marino2018}.
In a recent work, a semi-agnostic meta-model identical to ours was directly injected  in the analysis of the thermal emission from 7 qLMXB~\cite{Baillot2019}.
The constant radius approximation of Ref.~\cite{Guillot2013} was also performed with the new data, providing a radius of about $R_{NS}\approx11.06\pm0.4$~km.
Injecting constraints from nuclear physics and neglecting possible phase transitions in dense matter, the radius of a 1.4$M_\odot$ NS is predicted to be $R_{1.4}\approx12.4\pm0.4$~km.
The observation of a NS with a marked lower radius would clearly indicate a softening of the EoS induced by new degrees of freedom which are not contained in our nuclear physics meta-modeling.

For densities above $\sim 3\rho_{sat}$, new degrees of freedom could indeed appear, such as pion condensation~\cite{Haensel1982,Fenyi1992,Yasuda2018}, hyperonization~\cite{Colucci2013,Zdunik2013,Maslov2015,Chatterjee2016,Fortin2017,Gomes2019} or phase transition to quark matter~\cite{Han2013,Han2019,Montana2019,McLerran2019}.
In general the occurrence of new degrees of freedom tends to soften the EoS, and thus reduce the radius, except in the case of the quarkyonic model which describes the transition to quark matter as a crossover~\cite{McLerran2019}.
Since softening of the EoS also reduces the maximum mass it is important to include in the model selection the knowledge about $M_{max}^{obs}$.
The present analysis is focussed on nucleonic matter without phase transition, but we will show that it clearly calls for a future extension with phase transition(s).

The present work is organized as follows:
In Sec.~\ref{Theoretical framework}, the main theoretical inputs are presented, namely the nucleonic meta-model~\cite{Jerome2018p1},
and the general relativistic equations, i.e. the spherical Tollman, Oppenheimer, Volkoff (TOV)~\cite{Tolman1939,Oppenheimer1939,Book} and the pulsation equations~\cite{Hinderer2008,Hinderer2008v2,Damour2009}, which generate masses, radii and tidal deformabilities.
The statistical Bayesian tools are also introduced and we detail the construction of the posterior probability from the likelihood, which includes the constraints, and from the prior on the model parameters.
In Sec~\ref{Results and Discussions}, results are given: an analysis of the posterior PDF is undertaken for the following empirical parameters $L_{\textrm{sym}}$, $K_{\textrm{sym}}$, $Q_{\textrm{sat}}$ and $Q_{\textrm{sym}}$ as well as for the radius $R_{1.4}$ and the pressure at $2n_\textrm{sat}$, $P(2n_\textrm{sat})$.
Then the origin of the $L_\textrm{sym}$-$K_\textrm{sym}$ and $K_\textrm{sat}$-$Q_\textrm{sat}$ correlations are analyzed in details.
Finally we present our conclusions in Sec.\ref{Conclusions}.

\section{Theoretical framework}\label{Theoretical framework}

In the present study, we consider a few assumptions concerning the composition of the EoS.
We assume neutron stars cores are composed of neutrons and protons as well as a gas of electrons and muons at $\beta$ equilibrium.
No phase transition at high density is included here since we aim at exploring the limits of nucleonic hypothesis for the composition of the core of NS.

In addition, the nuclear model is requested to incorporate the bulk properties measured from experiments on finite nuclei, as well as to satisfy to the NS observations in terms of maximum observed NS mass and tidal deformability. The EoS should naturally satisfy causality and stability conditions~\cite{Rhoades1974} at all densities.

The link between NS matter and nuclear experiments can be performed through the nuclear empirical parameters, directly connected to the properties of the EoS.
These parameters are defined as the Taylor coefficients of the binding energy density for symmetric matter $e_{sat}$ and for the symmetry energy $e_{sym}$,
\begin{eqnarray}
e_{sat}(n_0) &=& E_{\textrm{sat}}+\frac{1}{2} K_{\textrm{sat}}x^2+\frac{1}{3!}Q_{\textrm{sat}}x^3 \nonumber \\
&&\hspace{1cm}+\frac{1}{4!}Z_{\textrm{sat}}x^4+O(x^5), \label{e3} \\
e_{sym}(n_0) &=& E_{\textrm{sym}}+L_{\textrm{sym}}x+\frac{1}{2}K_{\textrm{sym}}x^2+\frac{1}{3!}Q_{\textrm{sym}}x^3 \nonumber \\
&&\hspace{1cm}+\frac{1}{4!}Z_{\textrm{sym}}x^4+O(x^5). \label{e4}
\end{eqnarray}
where the Taylor expansion parameter is $x=(n_0-n_{sat})/(3n_{sat})$~\cite{Piekarewicz2016}, $n_0$ being the isoscalar density for protons and neutrons, $n_0=n_n+n_p$.
Assuming that these two quantities are the leading ones, the binding energy in asymmetric matter can be expressed as,
\begin{equation}\label{e2}
e(n_0,n_1)=e_{sat}(n_0) +\left( \frac{n_1}{n_0} \right)^2e_{sym}(n_0),
\end{equation}
where the isovector density is defined as $n_1=n_n-n_p$.
Note that Eq.~(\ref{e2}) neglects the small contribution beyond the quadratic terms in isospin asymmetry.

Completing this expansion with a kinetic energy term, a generic meta-model for nucleonic matter has recently been proposed and was tested to be able to reproduce most of existing nucleonic EoS~\cite{Jerome2018p1}. This recent approach will be considered in the following.

\subsection{The nuclear meta-modelling}\label{MM}

Given the assumptions previously listed, we consider a semi-agnostic approach which is mainly parametrized in terms of the nuclear empirical parameters (describing EoS fundamental properties such as the nuclear incompressibility) and can thus be easily related to experimental knowledge from nuclear physics.
At variance to fully agnostic approaches such as piece-wise polytropes~\cite{Raithel2017,Annala2018,Most2018,Fasano2019}
or sound speed model~\cite{Tews2018,Tews2019}, the meta-model can predict proton, electron and muons ratios as function of the density.
These ratios are controlled by the density dependence of the symmetry energy, and therefore the meta-model establish correlations between particle ratios and nuclear empirical parameters.
It allows to follow the $\beta$ equilibrium and any path out-of-equilibrium, such as the ones encountered in supernovae core collapse.
In the latter case, an extension at finite temperature is required, while here we compute the equation of state at zero temperature only.
For the sake of consistency, we briefly detail the main features of the meta-model. More details can be found in Ref.~\cite{Jerome2018p1}.

Since neutrons and protons are independent particles in the meta-model, the neutron and proton densities, $n_n$ and $n_p$, are defined as
\begin{equation}\label{e1}
n_{n/p}=\frac{1}{3\pi^2}{k^3_{F_{n/p}}} \, ,
\end{equation}
in terms of the Fermi momentum $k_{F_{n/p}}$.
From $n_n$ and $n_p$, one can define two equivalent quantities, which are
the isoscalar density ($n_0=n_n+n_p$) and the isovector density ($n_1=n_n-n_p$).
In the following, we will also use the density parameter $x=(n_0-n_{sat})/(3n_{sat})$ and the isospin asymmetry parameter $\delta=n_0/n_1$.
The two boundaries $\delta=0$ and 1 correspond to symmetric matter (SM) and to neutron matter (NM), respectively, while
any value of $\delta$ between -1 and 1 defines asymmetric nuclear matter.

In this work, we consider the metamodeling ELFc introduced in Ref.~\cite{Jerome2018p1}.
In this metamodeling, the energy per particle is defined as
\begin{equation}\label{e5}
 e(n_0,n_1)=t^{FG*}(n_0,n_1)+v(n_0,n_1).
\end{equation}
The first term is the kinetic energy density and the second term is the interaction potential.
The kinetic energy is related to the non-relativistic free Fermi gas (FG) as
\begin{eqnarray}
\nonumber t^{FG*}(n_0,n_1)&=& \frac{t^{FG}_{\textrm{sat}}}{2}\bigg(\frac{n_0}{n_{\textrm{sat}}}\bigg)^{2/3} \bigg[\bigg(1+\kappa_{\textrm{sat}}\frac{n_0}{n_{\textrm{sat}}}\bigg)f_1(\delta) \\
                          &&+\kappa_{\textrm{sym}}\frac{n_0}{n_{\textrm{sat}}}f_2(\delta) \bigg],    \label{e6}
\end{eqnarray}
where $t^{FG}_{\textrm{sat}}=3\hbar^2/(10m)(3\pi^2/2)^{2/3}n^{2/3}_\textrm{sat}$ is the kinetic energy per nucleons in SM and at saturation, $m$ is nucleonic mass taken identical for neutrons and protons $m=(m_n+m_p)/2=938.919\enspace\textrm{MeV}/c^2$, giving $t^{FG}_{\textrm{sat}}=22.1\enspace\textrm{MeV}$,
and the interaction potential can be expressed as
\begin{equation}\label{e7}
v(n_0,n_1)= \sum_{a \geq 0}^N {\frac{1}{a!}(c_a^{sat}+c_a^{sym}\delta^2)x^au_a(x)},
\end{equation}
where $u_a(x)=1-(-3x)^{N+1-a}$exp$(-bn_0/n_{\textrm{sat}})$ and $b$ is fixed to be $b=10$ln$2\approx6.93$.
In Eq.~(\ref{e6}), the functions $f_1$ and $f_2$ of the asymmetry parameter are defined as~\cite{Jerome2018p1}
\begin{eqnarray}
  f_1(\delta) &=& (1+\delta)^{5/3}+(1-\delta)^{5/3}, \label{e8} \\
  f_2(\delta) &=& \delta(1+\delta)^{5/3}-\delta(1-\delta)^{5/3}, \label{e9}
\end{eqnarray}
where
$f_1(\delta)$ represents an extension for isospin asymmetry and $f_2(\delta)$ includes the effect of Landau effective mass defined in Eq.~(\ref{e11}).
Besides,
the parameters $\kappa_{\textrm{sat}/\textrm{sym}}$ of Eq. (\ref{e6}) can be directly expressed in terms of the expected Landau effective mass at saturation density,
\begin{eqnarray}
\nonumber\kappa_{\textrm{sat}}=\frac{m}{m_{\textrm{sat}}^*}-1&=&\kappa_s,\thinspace\textrm{in SM}\thinspace(\delta=0), \\
  \kappa_{\textrm{sym}} = \frac{1}{2} \Bigg[\frac{m}{m^*_n}-\frac{m}{m^*_p}\Bigg]&=&\kappa_s-\kappa_v,\thinspace\textrm{in NM}\thinspace(\delta=1). \label{e11}
\end{eqnarray}
Fixing $\kappa_{\textrm{sat}/\textrm{sym}}$, the coefficients $c_a^{sat/sym}$ are directly related to the empirical parameters through the following one-to-one correspondences,
\begin{eqnarray}
\nonumber c_{a=0}^{sat} &=& E_{\textrm{sat}}-t_{\textrm{sat}}^{FG}(1+\kappa_{\textrm{sat}}), \\
\nonumber c_{a=1}^{sat} &=& -t_{\textrm{sat}}^{FG}(2+5\kappa_{\textrm{sat}}), \\
\nonumber c_{a=2}^{sat} &=& K_{\textrm{sat}}-2t_{\textrm{sat}}^{FG}(-1+5\kappa_{\textrm{sat}}), \\
\nonumber c_{a=3}^{sat} &=& Q_{\textrm{sat}}-2t_{\textrm{sat}}^{FG}(4-5\kappa_{\textrm{sat}}), \\
c_{a=4}^{sat} &=& Z_{\textrm{sat}}-8t_{\textrm{sat}}^{FG}(-7+5\kappa_{\textrm{sat}}), \label{e12}
\end{eqnarray}
and
\begin{eqnarray}
\nonumber c_{a=0}^{sym} &=& E_{\textrm{sym}}-\frac{5}{9}t_{\textrm{sat}}^{FG}[1+(\kappa_{\textrm{sat}}+3\kappa_{\textrm{sym}})], \\
\nonumber c_{a=1}^{sym} &=& L_{\textrm{sym}}-\frac{5}{9}t_{\textrm{sat}}^{FG}[2+5(\kappa_{\textrm{sat}}+3\kappa_{\textrm{sym}})], \\
\nonumber c_{a=2}^{sym} &=& K_{\textrm{sym}}-\frac{10}{9}t_{\textrm{sat}}^{FG}[-1+5(\kappa_{\textrm{sat}}+3\kappa_{\textrm{sym}})], \\
\nonumber c_{a=3}^{sym} &=&  Q_{\textrm{sym}}-\frac{10}{9}t_{\textrm{sat}}^{FG}[4-5(\kappa_{\textrm{sat}}+3\kappa_{\textrm{sym}})], \\
c_{a=4}^{sym} &=& Z_{\textrm{sym}}-\frac{40}{9}t_{\textrm{sat}}^{FG}[-7+5(\kappa_{\textrm{sat}}+3\kappa_{\textrm{sym}})]. \label{e13}
\end{eqnarray}

\begin{figure*}
\centering
\begin{subfigure}{0.49\textwidth}
\includegraphics[width=1\textwidth]{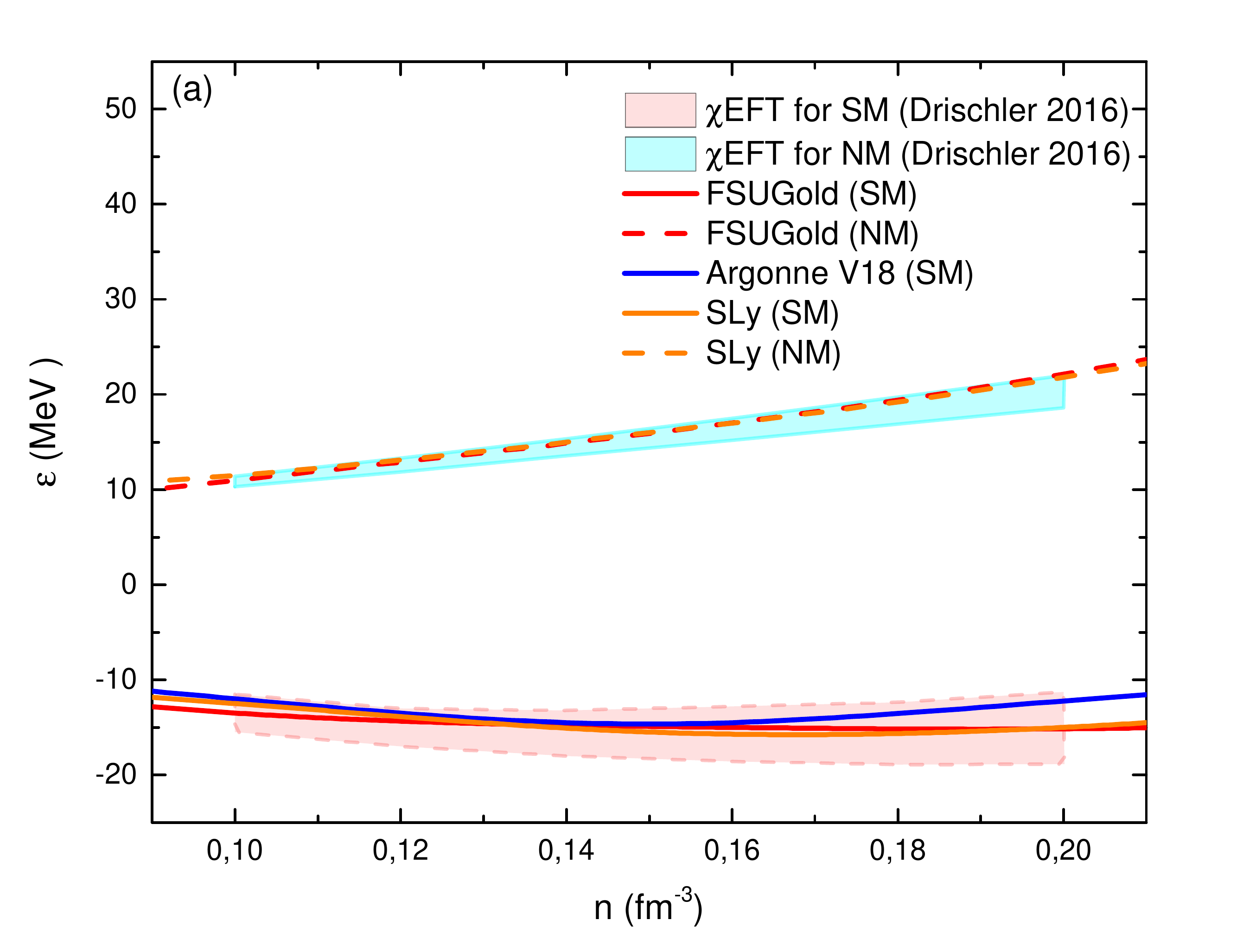}
\end{subfigure}
\begin{subfigure}{0.49\textwidth}
\includegraphics[width=1\textwidth]{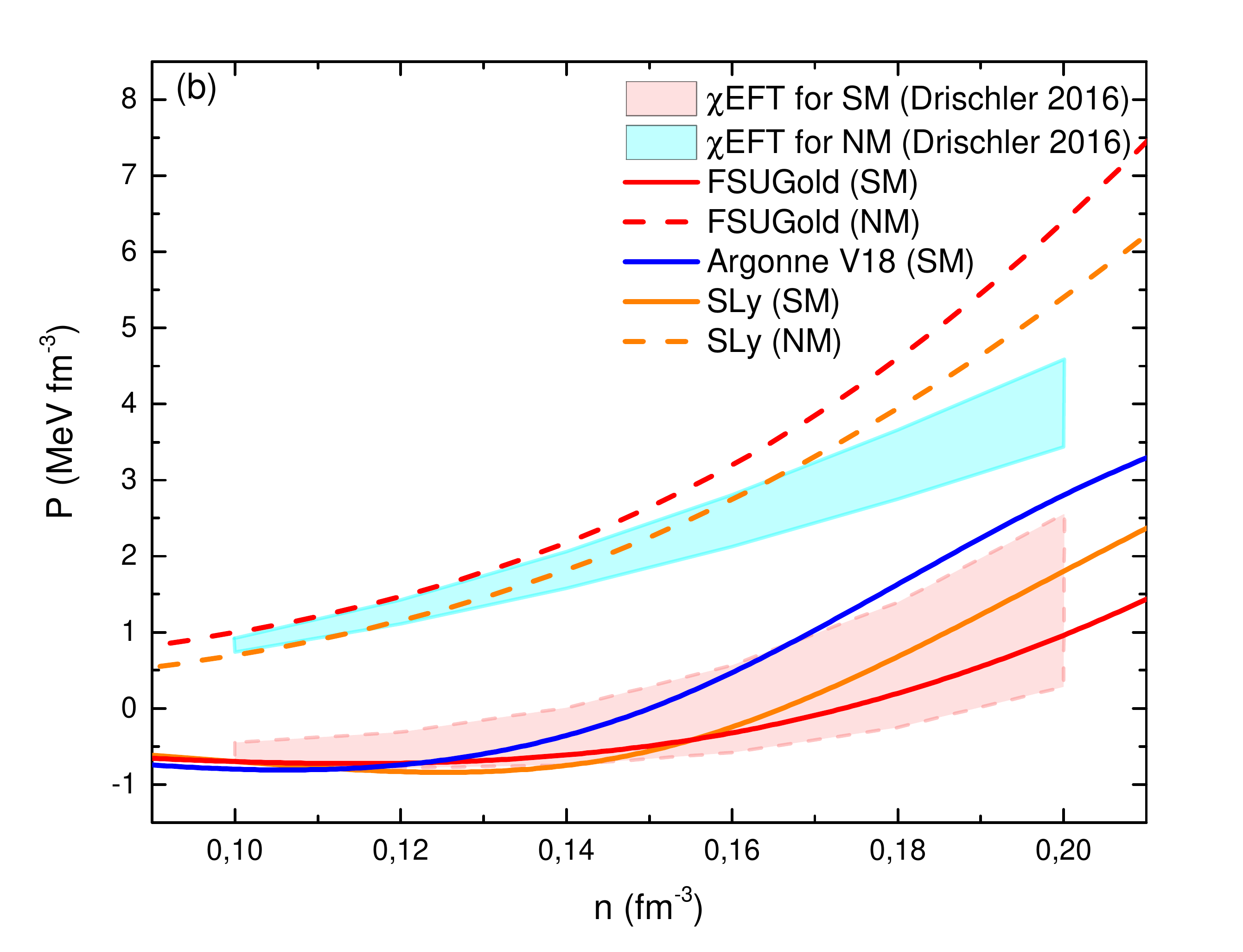}
\end{subfigure}
\caption{Energy (a) and Pressure (b) distributions calculated by using $\chi $EFT from the Ref.~\cite{Drischler2016} for both symmetric matter (SM) and neutron matter (NM).}
\label{f2}
 \end{figure*}

The one-to-one correspondence between the meta-model coefficients $c_a^{sat/sym}$ and the empirical parameters directly bridges the analysis of the impact of the empirical parameters on the properties of the equation of state and on the predictions for NS properties.
In the next subsection, we briefly detail how the NS properties such as masses, radii and tidal deformabilities can be related to the nuclear equation of state assuming general relativity (TOV and pulsation equation)~\cite{Book,Tolman1939,Oppenheimer1939,Hinderer2008,Hinderer2008v2,Damour2009}.

The advantage of the meta-model is that it is analytical, fast computed, very flexible and can reproduce most of existing nucleonic EoS. It keeps information concerning matter composition, such as the neutron/proton ratio, the fraction of electrons and muons. It is therefore optimal for extensive statistical analyses which require the set-up a large number of EoS samples.

At low densities, many-body perturbation theory based on $\chi$EFT nuclear two and three-body interactions have predicted bands based on 7 Hamiltonians which could equally well reproduce NN phase shifts and the binding energy of the deuteron~\cite{Drischler2016}.
These bands are represented in Fig.~\ref{f2} together with a set of models.
We compare these bands with three different models which are SLy~\cite{Chabanat1998}, ArgonneV18~\cite{Li2008} and FSUGold~\cite{Piekarewicz2009}.
The binding energies of these models are in good agreement with the $\chi$EFT bands in both symmetric matter (SM) and neutron matter (NM).
This is also true for the pressure in SM, but there are deviations in NM for FSUGold and SLy models, which predict a pressure above the bands for the high density region.
The origin of these deviation lies in the way the $\chi$EFT bands for the pressure is defined: It is the boundary calculated from the derivative of the binding energy predicted from the 7 Hamiltonians only. The pressure band does not exhaust all possible density dependence for the binding energy. It is therefore possible for models, such as FSUGold and SLy, to be inside the energy band and outside the pressure band.
The pressure band from the $\chi$EFT estimation provides a smaller band width than the one which would be based on all the models compatible with the energy band.
It is however the width compatible with the 7 Hamiltonians that we will consider in the following.

\subsection{Neutron star equilibrium properties and tidal deformability}\label{NSO}

The solution of the hydrostatic equations in general relativity for spherical and nonrotating stars, also named the Tollman, Oppenheimer, Volkoff (TOV) equations are expressed as~\cite{Book,Tolman1939,Oppenheimer1939};
\begin{eqnarray}
\nonumber  \frac{dm(r)}{dr} &=& 4\pi r^2\rho(r), \\
\nonumber  \frac{dP(r)}{dr} &=& -\rho c^2\Bigg(1+\frac{P}{\rho c^2} \Bigg)\frac{d\Phi(r)}{dr}, \\
  \frac{d\Phi(r)}{dr} &=& \frac{Gm}{c^2 r^2}\Bigg(1+\frac{4\pi P r^3}{m c^2} \Bigg)\Bigg(1-\frac{2Gm}{r c^2} \Bigg)^{-1}, \label{e14}
\end{eqnarray}
where $G$ is the gravitational constant, $c$ the speed of light in vacuum, $P$ the pressure, $m(r)$ the enclosed mass at radius r and
$\rho$ is the mass-energy density containing contributions from the nucleon rest mass ($m_N$) and from the total energy per particles ($e$): $\rho c^2=(m_N c^2+e)n_0$.

Since there are three equations for four variables ($m$, $P$, $\rho c^2$ and $\Phi$) in Eq.~(\ref{e14}), one need another equation to close the system.
This additional equation is provided by the equation of state of dense matter, $P(\rho c^2)$, which is evaluated at $\beta$-equilibrium for the NS conditions.
NSs are formed by a crust and a core whereas in its present form the meta-model only applies to uniform matter inside the core.
The core EoS is matched to the crust EoS with a cubic spline starting from an arbitrary transition density $n_{tr}=0.1n_{sat}$ to $n_\textrm{sat}$.
Below $n_{tr}$, we set crust EoS to be SLY for all core EoSs.
SLY is based on the Skyrme nuclear interaction SLy4~\cite{Chabanat1998}, which has been applied for the crust EOS considering a compressible liquid-drop model~\cite{Douchin2001}.
Besides we did not make an analysis for a crust EoS as well as $n_{tr}$, since we expect that the impact of the connection between the crust and the core is small for our analysis, for more details see Ref.~\cite{Jerome2018p2}.

These equations are defined in the Schwarzschild metric $ds^2=e^{2\Phi}c^2dt^2-e^{2\lambda}dr^2-r^2(d\theta^2+sin^2\theta d\phi^2).$
The potential $\Phi$ and the function $\lambda$ only depend on $r$, and the function $\lambda$ is fixed by $e^{-2\lambda}=1-2Gm/(c^2 r)$.
Eqs. (\ref{e14}) are integrated in coordinate space starting from $0$ to the radius $R$, fixing the boundary conditions $m(0)=0$ and $P(0)=P_c$ where $P_c(\rho=\rho_c)$ is arbitrarily varied.
The pressure $P$ decreases from the center to the surface and the NS radius is defined as the coordinate for which the condition $P(r=R)=0$ is reached.
The family of solutions with unique mass $m(R)=M$ and radii $R$ are generated by varying the central density $\rho_c$.

The tidal deformability $\Lambda$ resulting from the mutual interaction between two NS is defined as the quadratic metric perturbation in one NS generated as a response to the external field created by its companion.
The tidal deformability $\Lambda$ can be expressed in terms of the love number $k_2$ as~\cite{Hinderer2008,Hinderer2008v2}
\begin{equation}\label{e15}
 \Lambda=\frac{2 k_2}{3 C^5},
\end{equation}
where $C=(G M)/(c^2 R)$ is compactness of the NS for mass $M$ and radius $R$.
The love number $k_2$ is determined as,
\begin{eqnarray}
\nonumber  k_2&=& \frac{8C^5}{5}(1-2C)^2[2+2C(Y-1)-Y]  \\
\nonumber  &&\times  \big\{2C[6-3Y+3C(5Y-8)]           \\
\nonumber  &&+4C^3[13-11Y+C(3Y-2)+2C^2(1+Y)]            \\
\nonumber  &&+3(1-2C)^2[2-Y+2C(Y-1)]                      \\
  &&\times \textrm{ln}(1-2C)  \big\} ^{-1},      \label{e16}
\end{eqnarray}
where $Y=y(R)$ is the solution of pulsation equation at the surface of the NS.
The pulsation equation is expressed as~\cite{Hinderer2008,Hinderer2008v2};
\begin{equation}\label{e17}
r\frac{dy(r)}{dr}+y(r)^2+y(r)F(r)+Q(r)=0,
\end{equation}
with
\begin{eqnarray}
   F(r) &=& \frac{1}{r-2Gm/c^2} \Bigg(\frac{r+ 4\pi G r^3 }{P-\rho c^2}\Bigg), \label{e18}\\
\nonumber   Q(r)&=& \frac{4\pi G r^3/c^2}{r-2Gm/c^2}\Bigg(5\rho + \frac{9P}{c^2}+\frac{P+\rho c^2}{\rho c_s}\Bigg)\\
\nonumber   &&-\frac{4\pi G r^3/c^2}{r-2Gm/c^2}\Bigg(\frac{6}{4\pi G r^2/c^2} \Bigg)-\Bigg(\frac{2G^2 r}{c^4}\Bigg)\\
            &&\times \Bigg(\frac{m+4\pi r^3 P/c^2}{r-2Gm/c^2} \Bigg)^2, \label{e19}
\end{eqnarray}
where $c_s=dP/d\rho$ is the sound speed.
The pulsation equation is solved once the density and pressure radial profiles are defined from the solution of the TOV equations.

The wave-form extracted from the LIGO-Virgo GW interferometers is in fact impacted at the fifth-order by the two-NS combined
tidal deformability $\tilde{\Lambda}$, defined from each individual deformabilities of the NS, $\Lambda_1$ and $\Lambda_2$, as
\begin{equation}\label{e20}
\tilde{\Lambda} = \frac{16}{13}\frac{(M_1+12M_2)M_1^4\Lambda_1+(M_2+12M_1)M_2^4\Lambda_2}{(M_1+M_2)^5},
\end{equation}
where ($M_1$, $\Lambda_1$) and ($M_2$, $\Lambda_2$) are the masses and tidal deformabilities of the individual NSs (by convention $M_1\ge M_2$)~\cite{Ligo2017}.
If $M_1=M_2$, this expression becomes $\tilde{\Lambda}=\Lambda_1=\Lambda_2$.
However, as discussed below, we shall explore the asymmetric case in our study.

\subsection{Bayesian statistical analysis}\label{BA}

The relation between the empirical parameters and the NS properties is performed within the Bayesian statistical analysis.
The core of the Bayesian analysis lies on Bayes theorem expressing the probability associated to a given model, represented here by its parameters $\{a_i\}$, to reproduce a set of data, $P(\{a_i\} \mid \textrm{data})$ also called the posterior PDF, as~\cite{Book2}
\begin{equation}\label{e21}
 P(\{a_i\} \mid \textrm{data})\sim P(\textrm{data}\mid \{a_i\})\times P(\{a_i\}),
\end{equation}
where
$P(\textrm{data}\mid \{a_i\})$ is the likelihood function determined from the data comparison between the model and the measurement,
and $P(\{a_i\})$ is the prior PDF which represents our knowledge or bias on the model parameters.
Detailed discussions for the prior $P(\{a_i\})$ and for the likelihood probability $P(\textrm{data}\mid \{a_i\})$ are given in Secs.~\ref{M} and \ref{LHF}, respectively.

The marginal one- and two-parameter probabilities are defined as~\cite{Book2}
\begin{eqnarray}
  P(a_j \mid \textrm{data}) &=&\Bigg\{ \prod_{ \substack{i=1\\ i\neq j}}^5  \int da_i \Bigg\}  P(\{a_i\} \mid \textrm{data})\, ,  \label{e22a} \\
  P(a_j,a_k \mid \textrm{data}) &=& \Bigg\{ \prod_{ \substack{i=1\\ i\neq j,k }}^5  \int da_i \Bigg\}  P(\{a_i\} \mid \textrm{data})\, .  \label{e22b}
\end{eqnarray}
These marginal probabilities represent the one parameter PDF and the two-parameter correlation matrix, repectively.

\subsubsection{Fixed and varied parameters}

\label{M}
\begin{table}[t]
\centering
\tabcolsep=0.08cm
\def\arraystretch{1.5}
\begin{tabular}{ccccccc}
\hline\hline
 \begin{tabular}[c]{@{}l@{}}$E_\textrm{sat}$\\ (MeV)\end{tabular} & \begin{tabular}[c]{@{}l@{}}$E_\textrm{sym}$\\ (MeV)\end{tabular} & \begin{tabular}[c]{@{}l@{}}$n_\textrm{sat}$ \\ $(\textrm{fm}^{-3})$\end{tabular} & $m^*_\textrm{sat}/m$ & $\Delta m^*_\textrm{sat}/m$ &
 \begin{tabular}[c]{@{}l@{}}$Z_\textrm{sat}$ \\ $(\textrm{MeV})$\end{tabular}  &  \begin{tabular}[c]{@{}l@{}}$Z_\textrm{sym}$ \\ $(\textrm{MeV})$\end{tabular}  \\ \hline
 -15.8                                                            & 32.0                                                               & 0.155                                                                        & 0.75                         & 0.1 &0 & 0 \\
\hline\hline
\end{tabular}
\caption{The prior parameters: the fixed empirical parameters from group P1 and P3.}
\label{t1}
\end{table}

In our analysis, we evaluate the NS EOSs for each set of empirical parameters, which are 12 free parameters in total (10 empirical parameters and two parameters associated to the Landau effective mass).
Some of these parameters are however well-known and their small uncertainties do not impact the dense matter EoS to a large extend~\cite{Jerome2018p2}.
The 12 free parameters are therefore separated into three different groups:
\begin{itemize}
\item[(P1)] The parameters which are not varied: $E_\textrm{sat}$, $E_\textrm{sym}$, $n_\textrm{sat}$, $m^*_\textrm{sat}/m$ and $\Delta m^*_\textrm{sat}/m$.\\
\item[(P2)] The less-known parameters, which are varied on a uniform grid: $K_\textrm{sat}$, $L_\textrm{sym}$, $K_\textrm{sym}$, $Q_\textrm{sat}$ and $Q_\textrm{sym}$.
\item[(P3)] The totally unknown parameters, which however does not impact our analysis enough to be explored: $Z_\textrm{sat}$ and $Z_\textrm{sym}$.
\end{itemize}

In Table~\ref{t1}, we show the parameters which are not varied (from group P1), see Ref.~\cite{Jerome2018p2} and references therein.
The parameters like $E_\textrm{sat}$, $E_\textrm{sym}$ and $n_\textrm{sat}$ are well-known from finite-nuclei experiments and their uncertainty does not impact our analysis.
The other parameters such as $m^*_\textrm{sat}/m$ and $\Delta m^*_\textrm{sat}/m$ are also constrained from the nuclear experiments, to a lower extend, but their uncertainties only weakly impact dense matter EoS~\cite{Jerome2018p1}.

The varied parameters from the group P2 are discussed in the next paragraph.

The empirical parameters from the group P3 are fixed to be $Z_\textrm{sat/sym}=0$ since they do not play a major role for the dense matter equation of state associated to NS in the mass range between $1M_\odot$ and $2M_\odot$ which corresponds to possible masses of the binary NSs from GW170817 (see Ref.~\cite{Jerome2018p2} for more details).

\subsubsection{Discussion of the prior sets for the varied parameters}

In the present analysis the model parameters $\{a_i\}$ which are varied (group P2) are: $L_\textrm{sym}$, $K_\textrm{sat}$, $K_\textrm{sym}$, $Q_\textrm{sat}$ and $Q_\textrm{sym}$.
These empirical parameters are sampled on a uniform grid defined in Table~\ref{t2}.
These parameters are varied between a lower (Min) and an upper (Max) value, with N steps defining a constant step.
We have considered
two different choices for the prior.
In the prior set \#1, the boundaries of the parameters are determined such that the likelihood probability reaches zero, or a very small value compared to the one inside the range.
In the prior set \#2, we fix the boundaries to be the ones determined from nuclear physics experiments and reported in Ref.~\cite{Jerome2018p1}, except for $L_\textrm{sym}$ for which we allow the exploration of small values.
A detailed discussion about $L_\textrm{sym}$ is made in Sec.~\ref{Results and Discussions}.
The another point to fully explore $L_\textrm{sym}$ domain is choosing sufficient $K_\textrm{sym}$ values
since it contributes the pressure at densities $n>n_\textrm{sat}$ for $\beta$-equilibrated nuclear matter.
On the experimental side,
the maximum limit of $K_\textrm{sym}$ is unconstrained from the nuclear physics experiments.
However, the usual domain from nuclear models is $K_\textrm{sym}=100 \pm 100$ MeV~\cite{Jerome2018p1}.
Therefore, we chose the upper bound of $K_\textrm{sym}$ larger than the usual domain for both prior sets (see Ref.~\cite{Jerome2018p1} for more details about the complete analysis for $K_\textrm{sym}$).

\begin{table}[t]
\centering
\tabcolsep=0.15cm
\def\arraystretch{1.9}
\begin{tabular}{lcccccccccc}
\hline\hline
\begin{tabular}[c]{@{}c@{}}Empirical\\ Parameters\end{tabular} & \begin{tabular}[c]{@{}c@{}}$L_\textrm{sym}$\\ (MeV)\end{tabular} & \begin{tabular}[c]{@{}c@{}}$K_\textrm{sat}$\\ (MeV)\end{tabular} & \begin{tabular}[c]{@{}c@{}}$K_\textrm{sym}$\\ (MeV)\end{tabular} & \begin{tabular}[c]{@{}c@{}}$Q_\textrm{sat}$\\ (MeV)\end{tabular} & \begin{tabular}[c]{@{}c@{}}$Q_\textrm{sym}$\\ (MeV)\end{tabular} \\
\hline
\multicolumn{1}{l}{} & \multicolumn{5}{c}{Prior set \#1}    \\
Min                  & -10                                                              & \textbf{150}                                                              & -500                                                             & -1000                                                            & -2000 \\
Max                  & 70                                                               & 280                                                              & \textbf{1500}                                                             & 3000                                                             & 2000 \\
Step                 & 5                                                               & 10                                                               & 200                                                              & 400                                                              & 400 \\
N                    & 17                                                               & \textbf{14}                                                               & \textbf{11}                                                               & 11                                                               & 11 \\
\hline
\multicolumn{1}{l}{} & \multicolumn{5}{c}{Prior set \#2} \\
Min                  & -10                                                              & \textbf{180}                                                              & -500                                                             & -1000                                                            & -2000 \\
Max                  & 70                                                               & 280                                                              & \textbf{300}                                                             & 3000                                                             & 2000 \\
Step                 & 5                                                               & 10                                                               & 100                                                              & 400                                                              & 400 \\
N                    & 17                                                               & \textbf{11}                                                               & \textbf{9}                                                               & 11                                                               & 11 \\
\hline\hline
\end{tabular}
\caption{The prior parameters: the empirical parameters from group (P2), which are varied on a uniform grid for two different scenarios. Changes between the two sets are indicated in bold characters. Here
Min, Max are first and last values of the each parameter, Step is an increment for each iteration and N is the number of total fragment. For prior set \#1 and \#2,  see the text for details.}
\label{t2}
\end{table}

\subsubsection{Likelihood, error functions and filters}\label{LHF}

\begin{figure}
  \centering
  \includegraphics[width=0.5\textwidth]{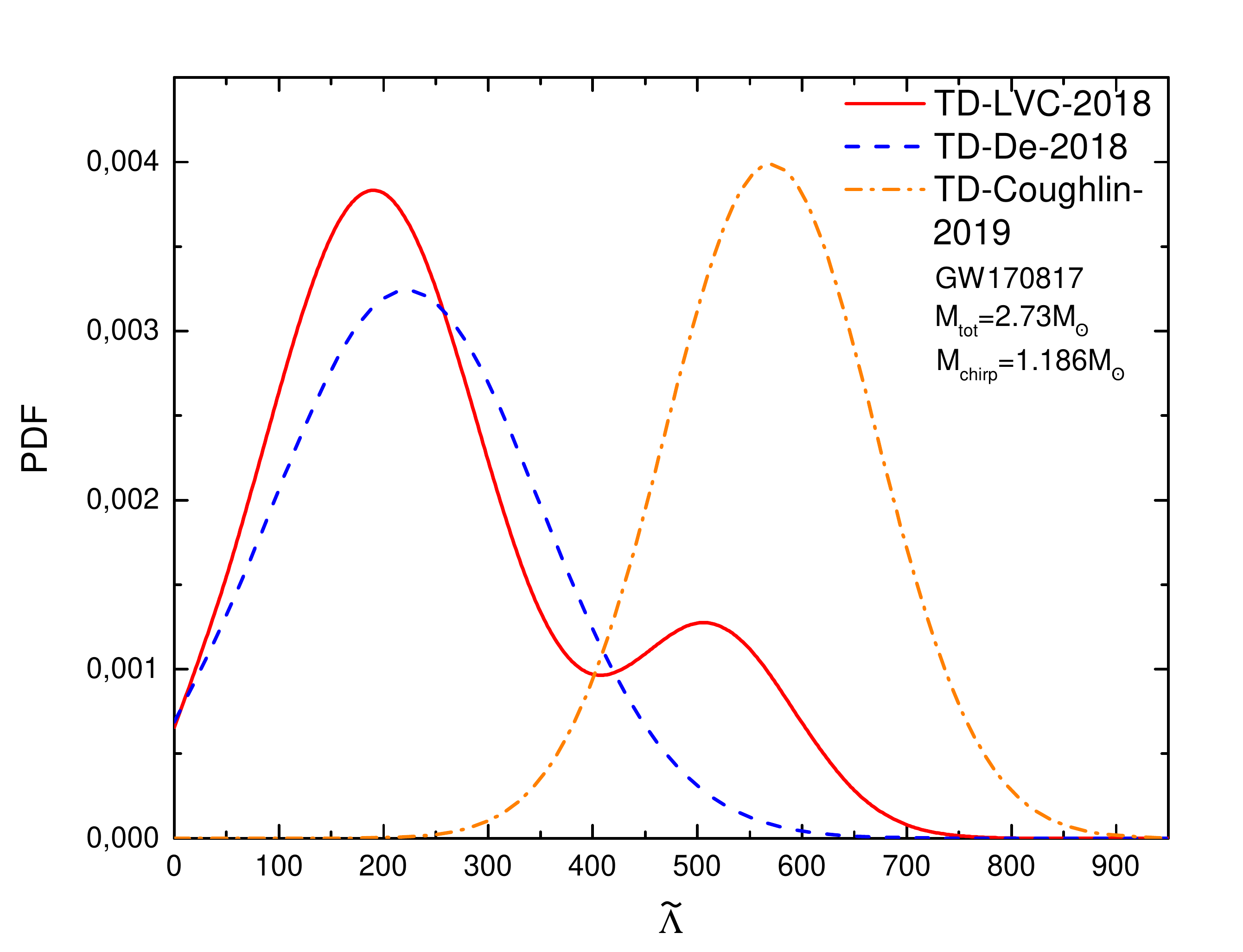}
  \caption{The tidal deformability PDF for various analyses of GW170817: TD-LVC-2018~\cite{Ligo2019}, TD-De-2018~\cite{De2018}, and TD-Coughlin-2019~\cite{Coughlin2019}.}
  \label{f1}
\end{figure}

The likelihood probability defines the ability of the model to reproduce the data. In the present analysis, it is defined as~\cite{Book2},
\begin{equation}\label{e23}
P(\textrm{data}\mid \{a_i\})=w_\textrm{filter} \times p_{\tilde{\Lambda}} \times p_\mathrm{\chi EFT} \times p_\mathrm{ISGMR}\, ,
\end{equation}
where $w_\textrm{filter}(\{a_i\})$ is a pass-band type filter which select only the models satisfying the necessary condition (C1) expressed hereafter,
and the probabilities $p_{\tilde{\Lambda}}$, $p_\mathrm{\chi EFT}$ and $p_\mathrm{ISGMR}$ are associated to constraints (C2)-(C4) expressed hereafter.
The constraints entering into the Bayesian probability (Eq. \ref{e23}) are:
\begin{itemize}
\item[(C1)] The necessary conditions that each viable EoS shall satisfy: causality, stability, positiveness of the symmetry energy and maximum observed mass $M_{max}^{obs}$.
\item[(C2)] $p_{\tilde{\Lambda}}$: the probability associated to the ability of the EoS to reproduce the tidal deformability extracted from the GW170817 event~\cite{Ligo2019,De2018,Coughlin2019}.
\item[(C3)] $p_\mathrm{\chi EFT}$: the probability measuring the compatibility between the meta-model and the energy and pressure bands function of the density  predicted from $\chi$EFT approach below saturation density~\cite{Drischler2016}.
\item[(C4)] $p_\mathrm{ISGMR}$: the probability of a given meta-model to be compatible with recent analysis of the ISGMR collective mode~\cite{Khan2012,Khan2013}.
\end{itemize}

The constraints (C1) are necessary constraints for all EoS, (C2) are constraints from astrophysics impacting high densities, while (C3) and (C4) are constraints from low-density nuclear physics. In the following, we detail how the probabilities associated to these constraints are estimated in practice.

Let us detail the constraints from the group (C1).
Causality, stability and positiveness of the symmetry energy are imposed as in Ref.~\cite{Jerome2018p2}.
The constraints are imposed up to the density corresponding to the maximum density of the stable branch.
We also impose that all viable EoS shall have a maximum mass $M_{max}\ge M_{max}^{obs}=2M_\odot$~\cite{Antoniadis2013}.

We now come to the constraint (C2) associated to the tidal deformability from GW170817.
We consider three independent GW analyses which provide different $\tilde{\Lambda}$ PDF.
These PDFs are displayed in Fig.~\ref{f1} under the caption "TD-LVC-2018", "TD-De-2018" and "TD-Coughlin-2019".
TD-LVC-2018 is the result of the latest analysis from the LIGO-Virgo collaboration~\cite{Ligo2019},
TD-De-2018 is an independent analysis proposed in Ref.~\cite{De2018} where more cycles has been considered and finally
TD-Coughlin-2019 is a recent analysis combining GW, EM and GRB signals in a Bayesian approach~\cite{Coughlin2019}.
Contrary to TD-De-2018 and TD-Coughlin-2019, TD-LVC-2018 has a double peak;
the highest one is peaked around $\tilde{\Lambda}_\textrm{max}^1 \approx 180$ and the smaller one is around $\tilde{\Lambda}_\textrm{max}^2\approx 550$.
However, in TD-De-2018, the only peak is $\tilde{\Lambda}_\textrm{max}\approx 200$ while in TD-Coughlin-2019 the peak is located close to the second one $\tilde{\Lambda}_\textrm{max}\approx 600$ .
The presence of a double peak has an impact on the $\tilde{\Lambda}$ range at 90\% confidence-level: the upper boundary is 720 in the case of TD-LVC-2018 while it is about 500 for TD-De-2018. The lower range is about 70 for TD-LVC-2018 and TD-Le-2018 while it is raised up to about 350 for TD-Coughlin-2019.
Anticipating our results, the PDF from TD-De-2018 select more compact objects than the others while the PDF from TD-Coughlin-2019 prefer less compact objects.

The probability $p_{\tilde{\Lambda}}$ is calculated in the following way.
For a given parameter set $\{a_i\}$, the TOV and the pulsation equations are first solved, which provides a family $\{M_i,\Lambda_i\}$, where $i$ is an index running over the central density. We then sample the mass distribution for the two NS ($M_1$, $M_2$) by taking a set of six masses, where $M_2$ is distributed from $1.1M_\odot$ to $1.35M_\odot$, and $M_1$ is calculated such that $M_1+M_2=2.73M_\odot$ accurately determined from GW170817. Note that eventually there are less masses in the sample if $M_1$ exceed the value $M_{max}$ for the EoS. For each sample elements the combined tidal deformability $\tilde{\Lambda}$ is calculated from Eq.~(\ref{e20}) and a probability, $p_{\tilde{\Lambda}}^k$, is assigned from the PDF shown in Fig.~\ref{f1} for the three scenarii. The final probability $p_{\tilde{\Lambda}}$ is then obtained from the averaging over the sample elements,
\begin{equation}\label{e24}
p_{\tilde{\Lambda}} = \frac 1 {N} \sum_{i=k}^{N} p_{\tilde{\Lambda}}^k \, .
\end{equation}

\begin{figure*}
\centering
\begin{subfigure}{0.49\textwidth}
\includegraphics[width=1\textwidth]{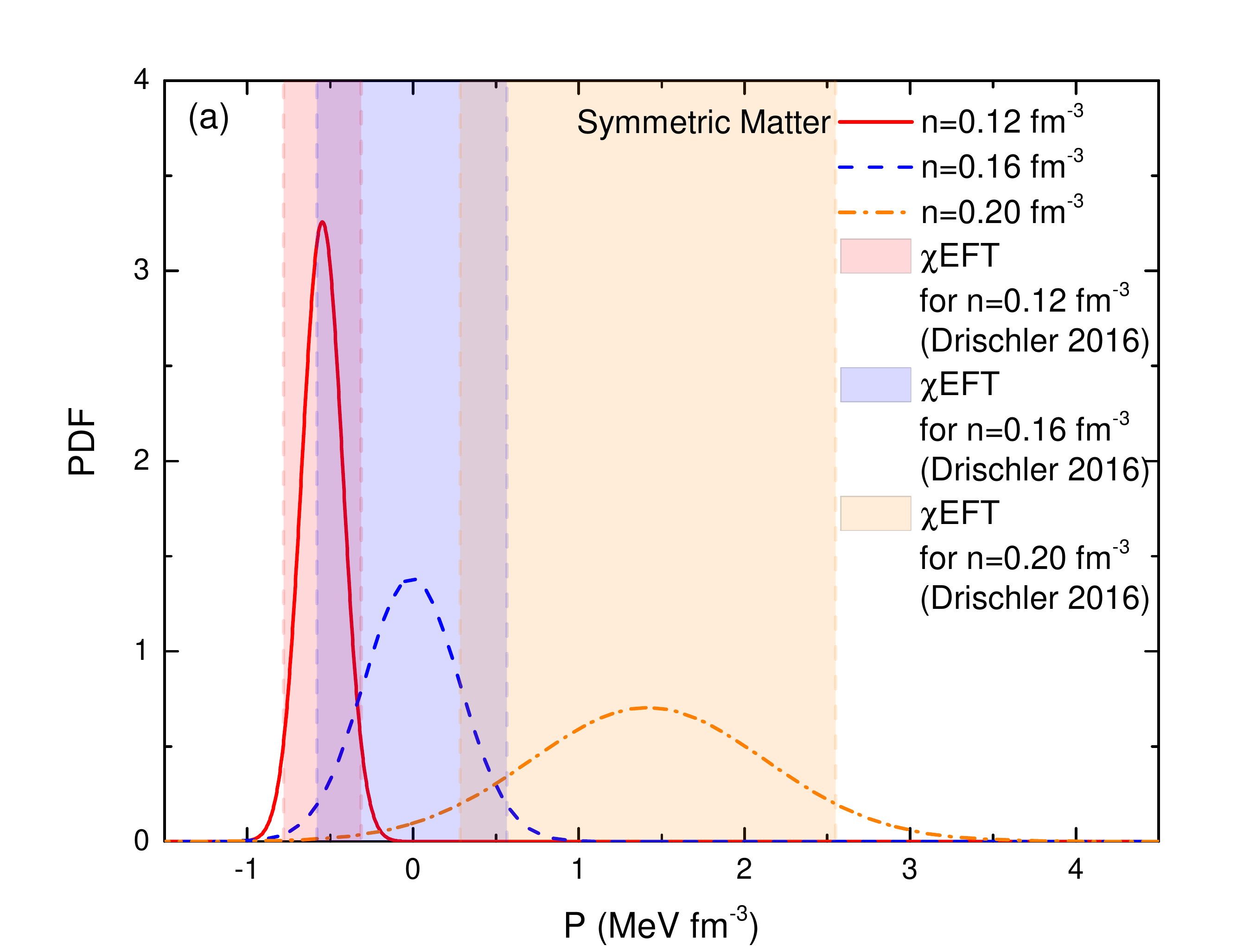}
\end{subfigure}
\begin{subfigure}{0.49\textwidth}
\includegraphics[width=1\textwidth]{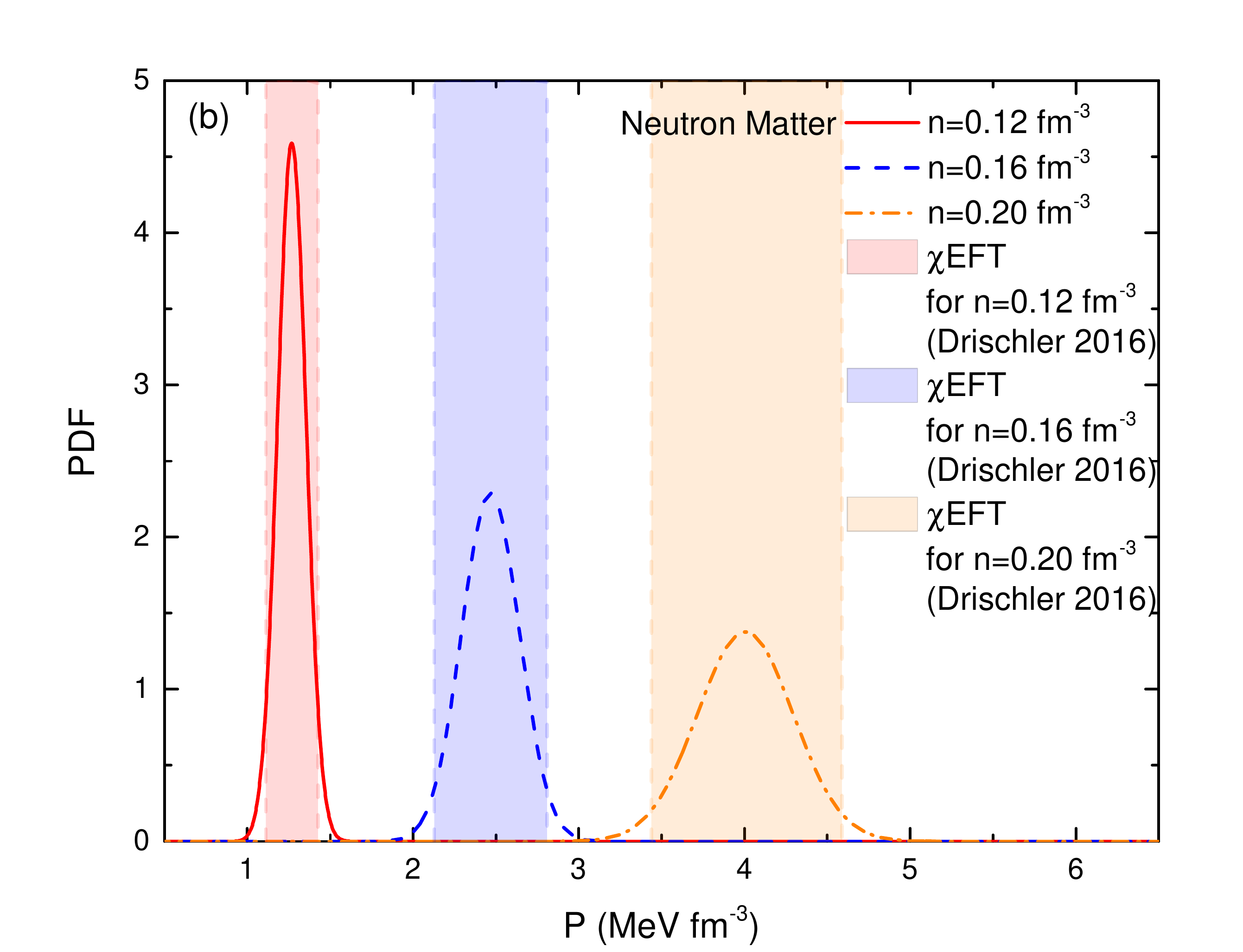}
\end{subfigure}
\caption{Pression posterior functions in neutron matter (NM) (a) and symmetric matter (SM) (b) obtained from the constraint C3 associated to the $\chi$EFT bands calculated in Ref.~\cite{Drischler2016}.}
\label{f3}
\end{figure*}

Note that there is no unique way to calculate $p_{\tilde{\Lambda}}$.
Another choice could have been, for instance, to assign to the parameter set the maximum probability obtained for $\tilde{\Lambda}$, $p_{\tilde{\Lambda}} = \max_{k} p_{\tilde{\Lambda}}^k$.
However, since the $\tilde{\Lambda}$ PDF only weakly depends on the mass asymmetry~\cite{De2018}, these two possible prescriptions are almost identical.
It should also be noted that in the case of a first-order phase transition occurring in the mass range under study, differences between these two prescriptions can be expected:
the mass asymmetry between the two NSs could have a strong impact on $\tilde{\Lambda}$ if the phase transition occurs at a mass in-between the ones of the two NS~\cite{Paschalidis2018,Tews2019}.


In the present analysis, we indeed assume that each neutron star of the binary system has the same EoS, the same particle composition and that their particle fractions is derived from the $\beta$ equilibrium condition.
Other exotic compositions such as pion or kaon condensation, Delta resonances, hyperons giving rise to Hybrid-Star/NS binaries will be considered in future works.

The constraint (C3) is a nuclear physics constraint which measures the proximity of the meta-model to the prediction bands for the energy per particle and the pressure in SM and NM obtained by many-body perturbation theory based on $\chi$EFT nuclear two and three-body interactions~\cite{Drischler2016}, see Fig.~\ref{f2} for illustration.
We remark that the considered $\chi$EFT band is acceptable with experimental results on finite nuclei (charge radius, neutron skin and also electric dipole polarizability and the weak form factor of $^{48}$Ca)~\cite{Hagen2016}.
Since it is calculated by using few-body observables at nucleonic scale with their theoretical uncertainties,
we can interpret (C3) as an common expectation of the nuclear physics.


In practice, we estimate the following error function $\chi_{2,\chi EFT}$ for each set of meta-models,
\begin{equation}\label{e25}
\chi_{2,\chi EFT}^2 = \frac 1 {N_{data}} \sum_{i=1}^{N_{data}} \left( \frac{o_i^{data}-o_i(\{a_i\})}{\sigma_i} \right)^2 \, .
\end{equation}
where $N_{data}=20$
is the number of data $o_i^{data}$ considered here, $o_i(\{a_i\})$ is the prediction of the model and $\sigma_i$ is associated to the uncertainties in the data and the accepted model dispersion.
We consider 5 density points
uniformly distributed between $0.12$~fm$^{-3}$ and $0.20$~fm$^{-3}$.
If $\Delta_i$ is the width of the band at each density point, we fix $\sigma_i=\Delta_i/2$ to ensure that 95\% of the models lie inside the band. The small tolerance of 5\% of the models outside the band is there to smoothly reduce the probability of marginal meta-models.
The associated probability is thus deduced from the usual Gaussian expression,
\begin{equation}\label{e26}
p_{\chi EFT} = \exp\left( {-\frac 1 2 \chi_{2,\chi EFT}} \right)\, .
\end{equation}

An example of likelihood function associated to the pression for a few densities (0.12, 0.16 and 0.20~fm$^{-3}$) is shown in Fig.~\ref{f3} for SM (a) and NM (b), where only the constraint C3 is imposed.
There is a nice overlap with all models inside $\chi$EFT bands with 95\% confidence level (shaded regions of Fig.~\ref{f3}).

The last constrain (C4) is obtained from a recent analysis of the ISGMR in finite nuclei~\cite{Khan2012,Khan2013}.
Theoretical models designed to describe finite nuclei and applied to the calculation of the ISGMR centroid energy in $^{120}$Sn and $^{208}$Pb suggest that the slope of the incompressibility $M_c$ at $n_c = 0.11\enspace \textrm{fm}^{-3}$ is very well correlated to the experimental data and less model-dependent than $K_{sat}$.
$M_c$ is defined as
\begin{equation}
M_c=3n_c \frac{dK(n_0)}{dn_0} \Bigg| _{n_0=n_c},
\end{equation}
where the incompressibility $K(n_0)$ in SM ($\delta=0$) is, $\chi$ being the compressibility,
\begin{equation}
K(n_0)= \frac{9 n_0}{\chi(n_0)}= 9n_{0}^2 \frac{d^2 e(n_0)}{dn_{0}^2} + \frac{18}{n_0}P(n_0) \, ,
\end{equation}
and the pressure is
\begin{equation}
P(n_0)=n_0^2 \frac{de(n_0)}{dn_0} \, ,
\end{equation}

It is found that $M_c = 1050\enspace\textrm{MeV}\pm  50\enspace\textrm{MeV}$~\cite{Khan2012,Khan2013}.
The interesting feature of this parameter is that it is much less model dependent that the more frequently considered incompressibility modulus $K_{sat}=K(n_{sat})$.

In practice, we calculate the value of $M_c$ for each of our meta-models by assigning the following probability,
\begin{equation}\label{e30}
p_{ISGMR} = \exp\left\{ -\frac 1 2 \left( \frac{M_c(\{a_i\})-1050}{25}\right)^2  \right\}\, ,
\end{equation}
where we associate the dispersion $\pm 50$~MeV estimated in Refs.~\cite{Khan2012,Khan2013} to the distribution of 95\% of the meta-models.


\section{Results and Discussions}
\label{Results and Discussions}

Taking advantage of the Bayesian framework, we analyse the contributions coming from the constraints (C2)-(C4) to understand the individual contributions coming from $\tilde{\Lambda}$, $\chi$EFT and ISGMR to the final posterior probability.
Both joint and sole posterior probabilities will be shown, and the influence of the prior set and three $p_{\tilde{\Lambda}}$ are also presented.
In the following, the uncertainties are defined as the 68\% confidence level around the centroid.

In the present statistical analysis, we generate a sample of $294\ 151$ parameter sets for prior set \#1 and $203\ 643$ for prior set \#2 before the filtering (see Table~\ref{t2}).
For each set, the probabilities $p_{\tilde{\Lambda}}$, $p_\mathrm{\chi EFT}$ and $p_\mathrm{ISGMR}$ are calculated according to Eqs.~(\ref{e24}), (\ref{e26}) and (\ref{e30}).
The total likelihood probability is calculated from Eq.~(\ref{e23}).
The reduction from the multi-dimension PDF to the one- or two-parameter probabilities are obtained from marginalization, see Eqs.~(\ref{e22a}) and (\ref{e22b}).

In the present section, we analyze the PDF for $L_\textrm{sym}$, $K_\textrm{sym}$, $Q_\textrm{sat}$, $Q_\textrm{sym}$, $R_{1.4}$, $P(2n_\textrm{sat})$ and the correlations between the parameters $L_\textrm{sym}$-$K_\textrm{sym}$ and $K_\textrm{sat}$-$Q_\textrm{sat}$.
The PDF for $K_{sat}$ is not shown here since $K_{sat}$ is found to only have a weak impact on $p_{\tilde{\Lambda}}$.

\subsection{Probability distributions for the empirical parameters} \label{EP}

We first study posterior distribution for the empirical parameters: $L_\textrm{sym}$,  
$K_\textrm{sym}$, $Q_\textrm{sat}$, $Q_\textrm{sym}$.

\subsubsection{Empirical parameter $L_\textrm{sym}$}

\begin{figure}
\centering
\begin{subfigure}{0.5\textwidth}
\includegraphics[width=1\textwidth]{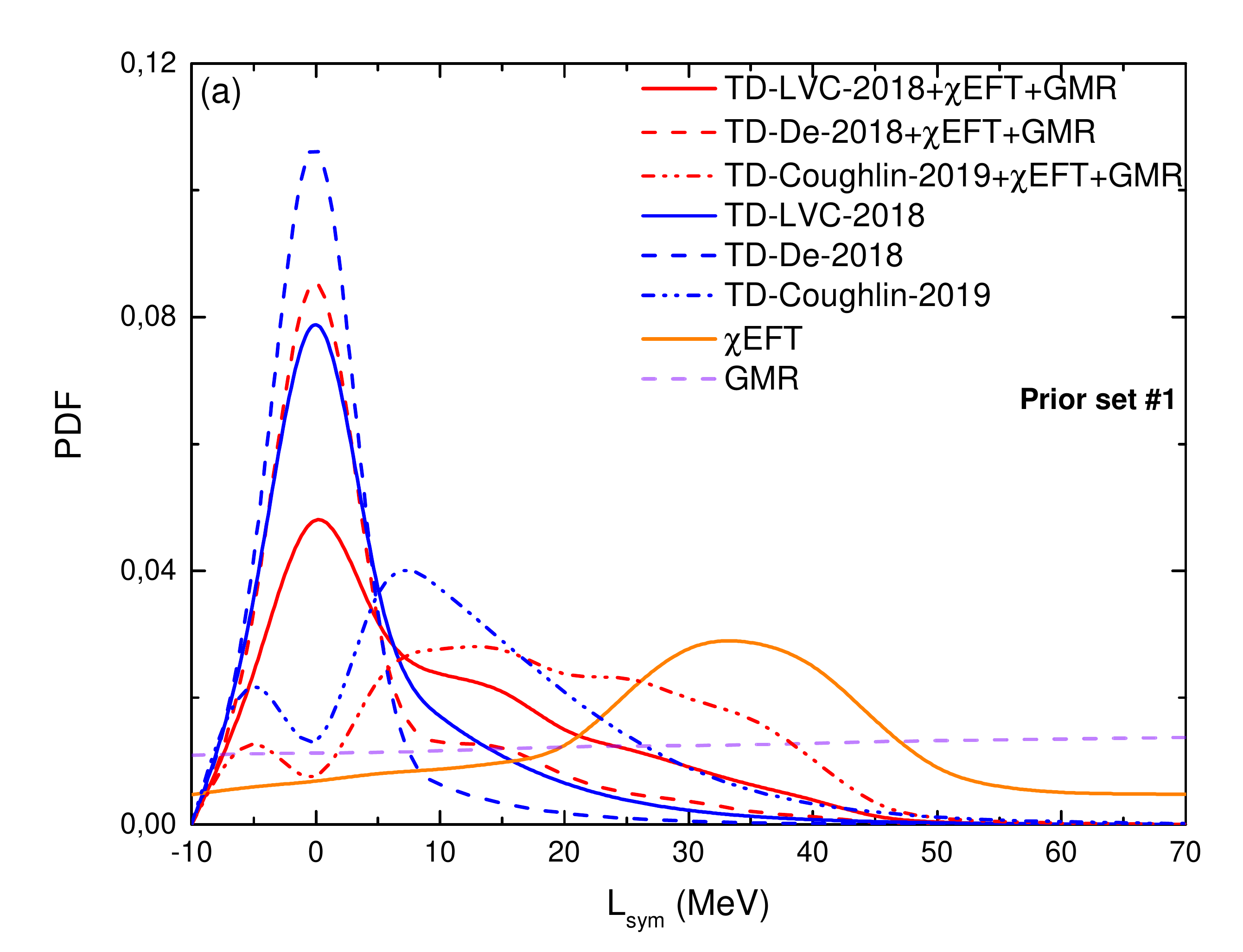}
\end{subfigure}
\begin{subfigure}{0.5\textwidth}
\includegraphics[width=1\textwidth]{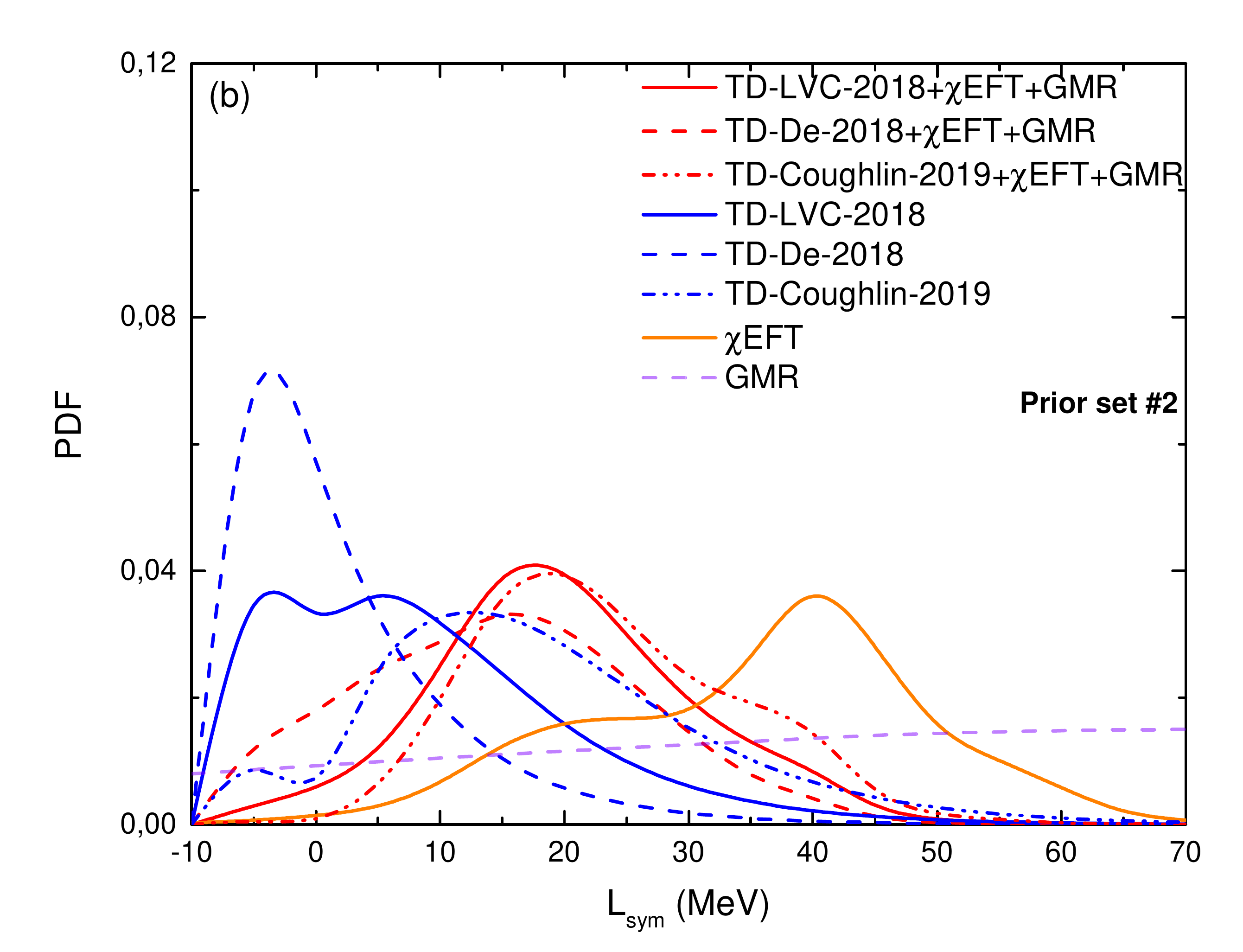}
\end{subfigure}
\caption{The generated PDFs of $L_\textrm{sym}$ for the prior set \#1 (a) and the prior set \#2 (b).}
\label{f5}
\end{figure}

The empirical parameter $L_\textrm{sym}$ is the slope of the symmetry energy at $n_\textrm{sat}$.
In Fig.~\ref{f5} the detailed contributions of the constraints (C2)-(C4) as well as of the role of the $\tilde{\Lambda}$-PDF and of the prior scenario \#1 (panel a) or \#2 (panel b) is shown.
Note the marked tension between the PDF associated to $\chi$EFT and the $\tilde{\Lambda}$ one (TD-LVC-2018, TD-Le-2018, TD-Coughlin-2019).
Being peaked at higher values for $\tilde{\Lambda}$, the TD-Coughlin-2019 PDF favors slightly larger $L_\textrm{sym}$ values than the two others.
The influence of the prior is weak, but interestingly, the prior set \#1 produces more peaked posteriors than the prior set \#2, which is inferred from analyses of nuclear physics models.
This could be interpreted as a signal for the marked deviations from nuclear physics predictions: when the constraints from nuclear physics is relaxed (mainly the prior on $K_\textrm{sym}$) in the set \#1, there is a group of EoS which are clearly preferred by the GW tidal deformability and which are located well outside the domain for $L_\textrm{sym}$ suggested by nuclear physics.

The GMR constraint has no effect on $L_\textrm{sym}$ since the GMR mainly contributes to parameters related to symmetric nuclear matter.
The $\chi$EFT constraint predicts values for
$L_\textrm{sym}=35^{+7}_{-10}/42^{+7}_{-16}$~MeV for the prior set \#1/\#2,
while the tidal deformability favors low or even negative $L_\textrm{sym}$ values.
For instance, TD-LVC-2018 gives
$L_\textrm{sym}=0^{+5}_{-3}/-3^{+18}_{-3}$~MeV for the prior sets \#1/\#2.
As expected, the prior set \#2 allows some positive values for $L_\textrm{sym}$ in the PDF shown in Fig.~\ref{f5}.

The joint probabilities naturally favor values for $L_\textrm{sym}$ which are intermediate between the two extremes.
The most probable value for TD-LVC-2018 (TD-De-2018 and TD-Coughlin-2019) is
$L_\textrm{sym}=0^{+12}_{-4}/17^{+11}_{-7}$~MeV
($L_\textrm{sym}=0^{+2}_{-3}/15^{+11}_{-13}$~MeV and $L_\textrm{sym}=10^{+7}_{-10}/16^{+15}_{-5}$~MeV) for the prior set \#1/\#2.
The difference between the prior sets \#1 (panel a) and \#2 (panel b) reflects the choice for the prior distribution:
the upper bound for $K_\textrm{sym}$ is fixed to be 1500~MeV for the prior set \#1 and only 300~MeV  for the prior set \#2 (see Table~\ref{t2}).
The distribution of $L_\textrm{sym}$ is thus impacted by the knowledge from the next order empirical parameter $K_\textrm{sym}$: The better defined $K_\textrm{sym}$, the more peaked $L_\textrm{sym}$.
The correlation between $L_\textrm{sym}$ and $K_\textrm{sym}$ will be analysed in Sec.~\ref{Cor}.
Note that the influence of the unknown high order empirical parameters was originally stressed in Ref.~\cite{Jerome2019}.

Interestingly, the empirical parameter $L_\textrm{sym}$ is investigated by a large number of experiments, see Ref.~\cite{Li2013} and references therein.
Confronting the predictions of various nuclear physics experiments, namely neutron skin thickness, heavy ion collisions, dipole polarizability, nuclear masses, giant dipole resonances and isobaric analog states, the values of $L_\textrm{sym}$ vary between $30$ and $70$ MeV~\cite{Li2013,Jerome2018p1,Maza2018,De2018}.
It is however interesting to note that a few studies give for $L_\textrm{sym}$ lower values, even negative ones, see Refs.~\cite{Berdichevsky1985,Berdichevsky1988}, from the charge radius of Sn and Pb isotopes using a droplet model.
A detailed analysis based on a few Skyrme and Gogny interactions advocates also for a low values for $L_\textrm{sym}$~\cite{Baldo2016}.
The measurement of the $^{208}\textrm{Pb}$ neutron skin thickness from the PREX collaboration (Lead Radius Experiment~\cite{PREX1}) is expected to provide a model independent estimation of $L_\textrm{sym}$.
The experiment has however not yet been very conclusive, with a measured neutron skin thickness $R_\textrm{skin}^\textrm{208}=0.33^{+0.16}_{-0.18}$~fm.

Anticipating the results of Sec.~\ref{RP}, there is a strong correlation between the marginalized probability distribution as function of $L_\textrm{sym}$ and the one as function of $R_{1.4}$: a low value of $L_\textrm{sym}$ coincides with a low radius  $R_{1.4}$.
Hence the peak at low $L_\textrm{sym}$ observed for the tidal deformabilities TD-LVC-2018 and TD-De-2018 reflects that the $\tilde{\Lambda}$ PDF prefer NS with small radii.
Since the physical implications are more clear in terms of radii, we further discuss the meaning of low radii (equivalently low $L_\textrm{sym}$) in Sec.~\ref{RP}.

\subsubsection{Empirical parameter $K_\textrm{sym}$}

\begin{figure}
\centering
\begin{subfigure}{0.5\textwidth}
\includegraphics[width=1\textwidth]{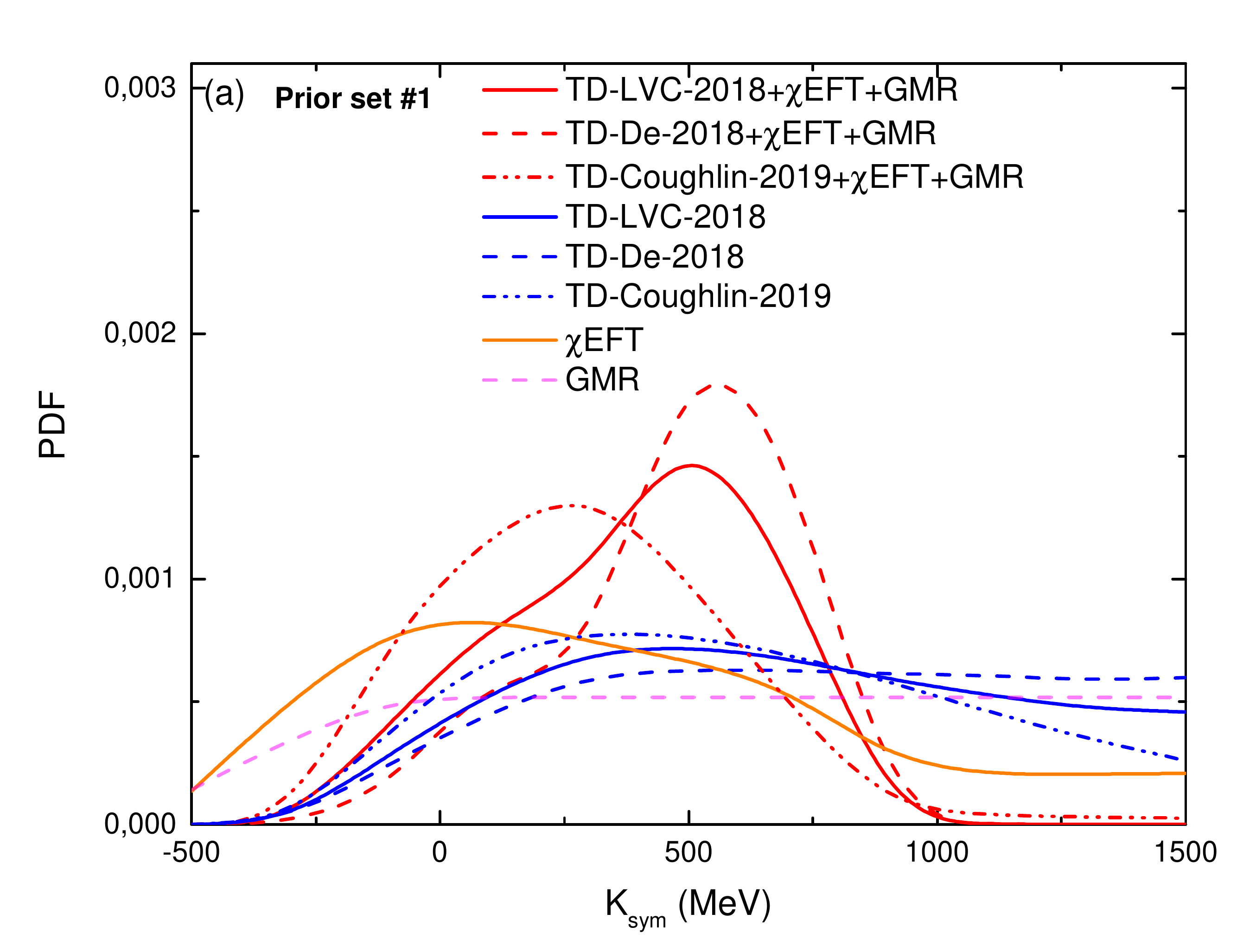}
\end{subfigure}
\begin{subfigure}{0.5\textwidth}
\includegraphics[width=1\textwidth]{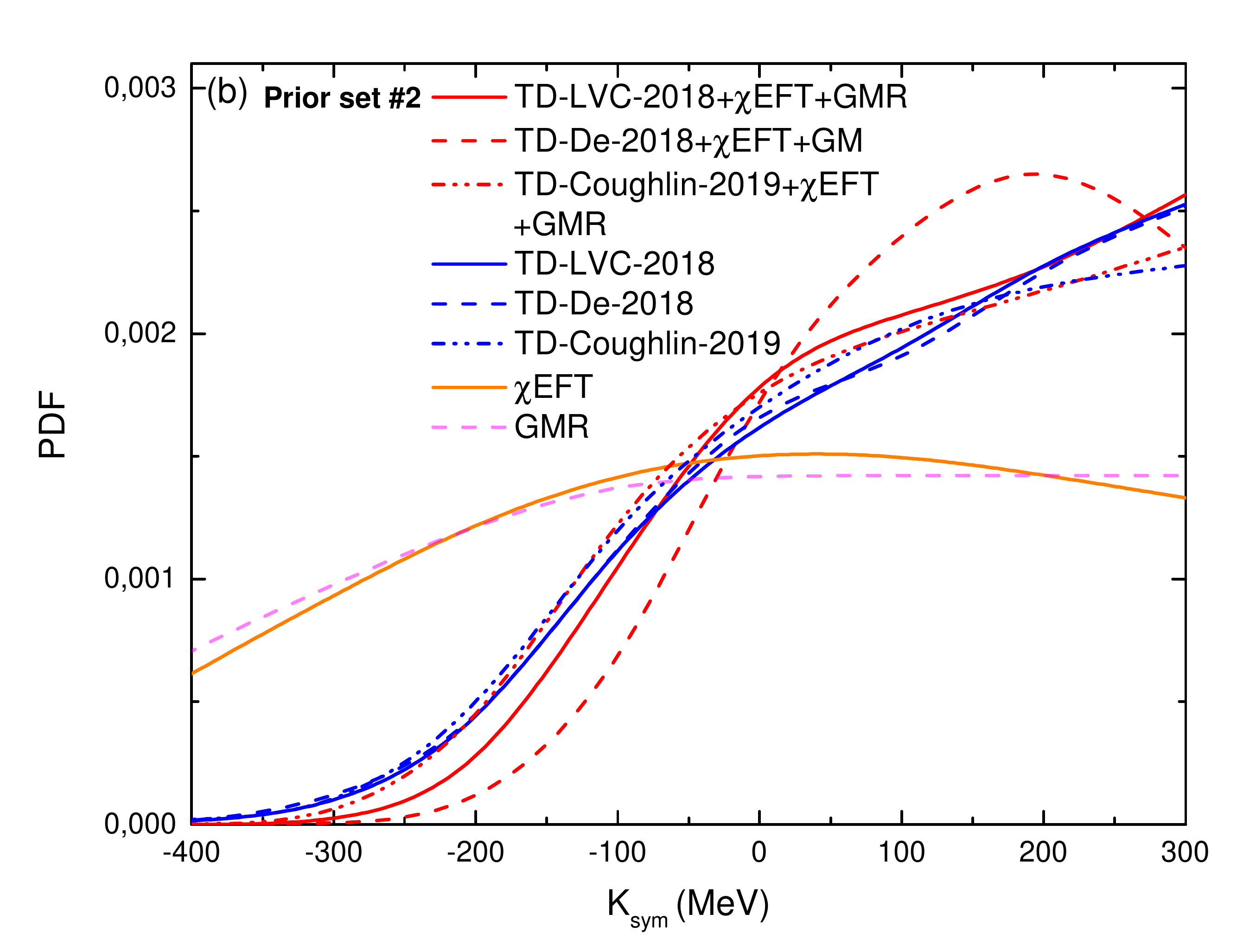}
\end{subfigure}
\caption{The generated PDFs of $K_\textrm{sym}$ for the prior set \#1 (a) and \#2 (b).}
\label{f6}
\end{figure}

The empirical parameter $K_\textrm{sym}$ encodes the curvature of the symmetry energy at $n_\textrm{sat}$.
It is different from the parameter $K_\tau$ which is defined as the curvature of the binding energy for a fixed proton fraction~\cite{Piekarewicz2009},
\begin{equation}
K_\tau = K_\textrm{sym} - 6 L_\textrm{sym}- Q_\textrm{sat} L_\textrm{sym}/K_\textrm{sat} \, .
\label{e31}
\end{equation}
The isospin dependence of the isoscalar giant monopole resonance (ISGMR) is a natural observable to determine the parameter $K_\tau$~\cite{Piekarewicz2009}. $K_\tau=-550\pm100$~MeV has been extracted from the breathing mode of Sn isotopes (Refs.~\cite{Li2007,Garg2007}) and also from isospin
diffusion observables in nuclear reactions (Refs.~\cite{Li2005,Li2008v2}).
If $L_\textrm{sym}$ and $Q_\textrm{sat}$ were well determined, Eq.~(\ref{e31}) would provide an equivalence between $K_\tau$ and $K_\textrm{sym}$.
However, the large uncertainties on $L_\textrm{sym}$ and $Q_\textrm{sat}$ induce a large error bar for $K_\textrm{sym}$, of the order of $\pm 600$~ MeV~\cite{Jerome2018p1}.
Besides, the statistical analysis of various theoretical model predict a value $K_\textrm{sym}=-100\pm 100$~MeV~\cite{Jerome2018p1}.
This result is also in agreement with Ref.~\cite{Carson2019}, which is GW analysis done by using TE EoSs.
On the other hand,
there is an experimental determination of $K_\textrm{sym}$ by using latest ISGMR values of $^{90}$Zr, $^{116}$Sn and $^{208}$Pb nuclei from Skyrme EDFs: $K_\textrm{sym}=-120\pm40$~MeV from Ref.~\cite{Sagawa2019}.
The smaller error bar than the statistical analysis reveals the presence of correlations between $L_\textrm{sym}$, $Q_\textrm{sat}$ and $K_\textrm{sym}$ which does not vary independently from each other.

In our analysis, we explore two priors for $K_\textrm{sym}$, one which is pushed until the likelihood probability is quenched (prior set \#1), and one which is compatible with the expectation  $K_\textrm{sym}=-100\pm 100$~MeV (prior set \#2).
In Fig.~\ref{f6}, the posterior PDFs for $K_\textrm{sym}$ are displayed for both prior sets.
The posteriors are qualitatively similar between the prior sets \#1 and \#2.
From $\chi$EFT, we obtain
$K_\textrm{sym}=15^{+600}_{-265}/10^{+290}_{-410}$~MeV for the prior set \#1/\#2.
The tidal deformability however favors positive values where TD-LVC-2018 (TD-De-2018 and TD-Coughlin-2019) predicts
$K_\textrm{sym}=375^{+\infty}_{-400}$~MeV ($K_\textrm{sym}=390^{+\infty}_{-400}$ and $K_\textrm{sym}=275^{+890}_{-330}$~MeV) for the prior set \#1.
TD-Coughlin-2019 prefer values for $K_\textrm{sym}$ very slightly below the distributions produced by TD-LVC-2018 and TD-De-2018.
This can be understood from the $L_\textrm{sym}$-$K_\textrm{sym}$ anti-correlation originating in the causality condition.
Although we cannot define centroid values of $K_\textrm{sym}$ since the prior set \#2 limits the posteriors to $K_\textrm{sym}=300$~MeV,
shifting the prior set \#1 to \#2 adds 100~MeV to the minimum values of $K_\textrm{sym}$.
There is also a difference between the expectations from $\chi$EFT and from the tidal deformability, while at variance with $L_\textrm{sym}$, the differences are here less marked.
The impact of the ISGMR is also pretty small.

Finally, the joint probabilities shown in Fig.~\ref{f6} give
$K_\textrm{sym}=440^{+210}_{-210}$~MeV ($K_\textrm{sym}=560^{+150}_{-150}$~MeV and $K_\textrm{sym}=260^{+240}_{-240}$~MeV) for TD-LVC-2018 (TD-De-2018 and TD-Coughlin-2019).
Considering the $-2 \sigma_\textrm{min}$ value for each centroid,
one can define the lower limit for $K_\textrm{sym}$: $K_\textrm{sym}\geq 18$~MeV for TD-LVC-2018,
$K_\textrm{sym}\geq 260$~MeV for TD-De-2018 and $K_\textrm{sym}\geq -213$~MeV for TD-Coughlin-2019.
It should be noted that several analysis have been done on the bounds of $K_\textrm{sym}$, providing $K_\textrm{sym}\geq-500\enspace\textrm{MeV}$ to $K_\textrm{sym}\geq-250\enspace\textrm{MeV}$ depending on used models~\cite{Mondal2017,Chen2009,Yoshida2006,Danielewicz2009}.
Besides, an interesting work about the lower limit of $K_\textrm{sym}$ is the Unitary Gas (UG) limit for the NM, which is in a good agreement with our predictions~\cite{Tews2017b}.
Since the ground state energy per particle in the UG is proportional to the Fermi energy, one can describe a forbidden zone for energy per particle of EoS in terms of the Fermi energy for neutron matter.
Using the average value of $K_\textrm{sat}=230\pm20\enspace\textrm{MeV}$ (see Ref.~\cite{Jerome2018p1} for a complete analysis about the parameter $K_\textrm{sat}$),
a minimum limit for $K_\textrm{sym}$ can be obtained: $K_\textrm{sym}\geq-255\pm20\enspace\textrm{MeV}$.
However, contrary to the UG, the NM includes effective-range effects and interactions in higher partial waves especially for densities $n\geq n_\textrm{sat}$.
Therefore, it is expected that the lower limit of $K_\textrm{sym}$ should be higher then the one obtained from the UG.

\subsubsection{Empirical parameter $Q_\textrm{sat}$}

\begin{figure}
\centering
\begin{subfigure}{0.5\textwidth}
\includegraphics[width=1\textwidth]{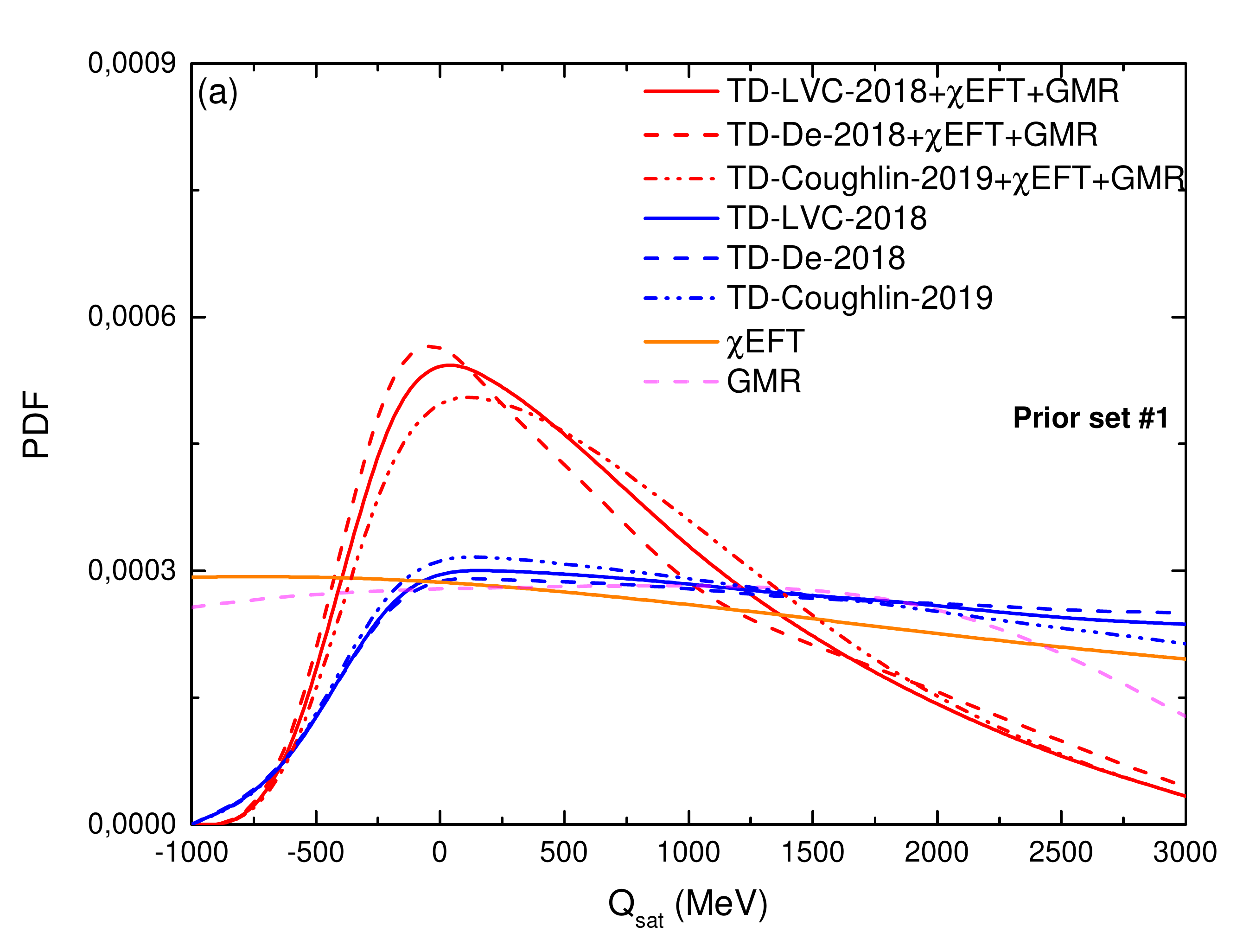}
\end{subfigure}
\begin{subfigure}{0.5\textwidth}
\includegraphics[width=1\textwidth]{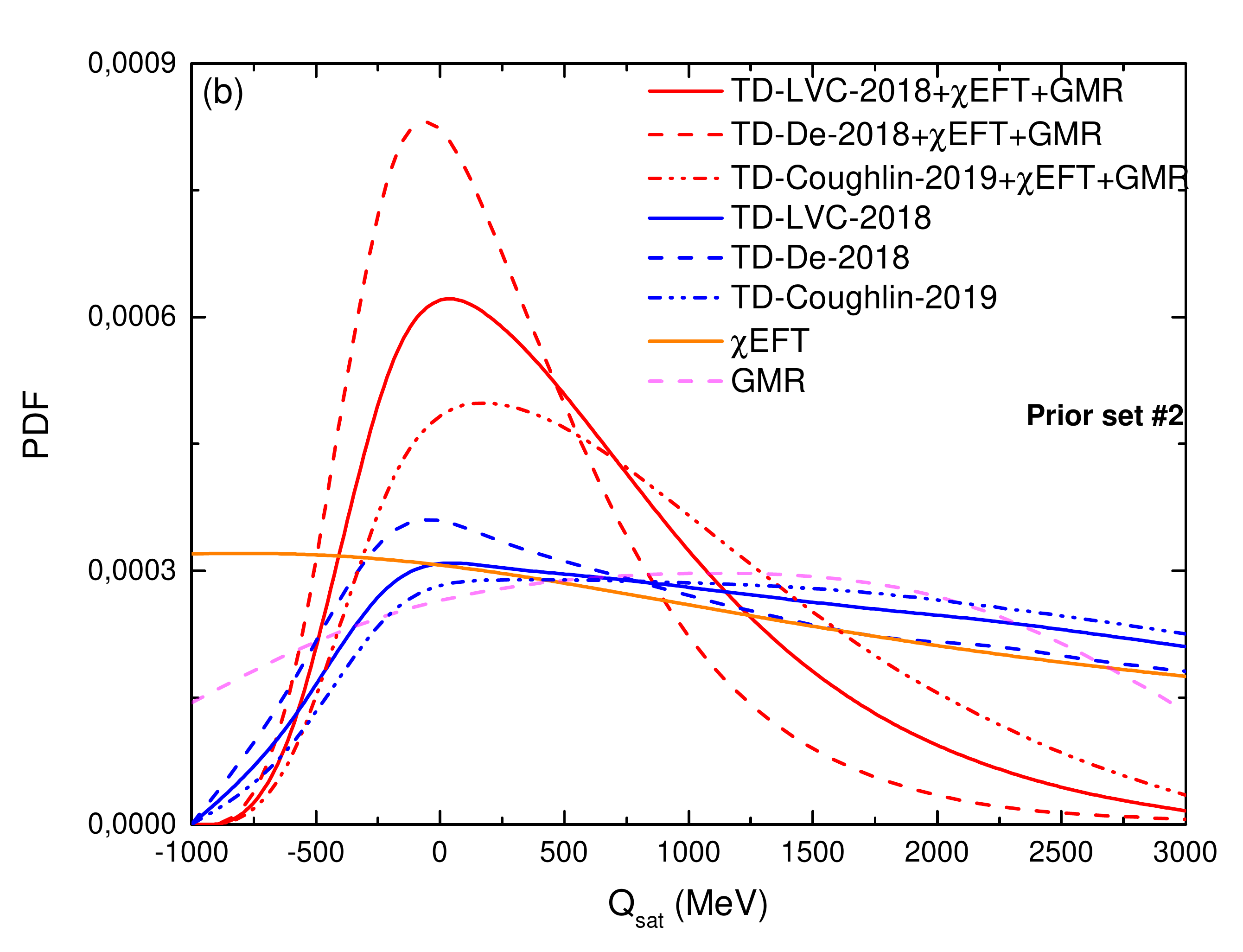}
\end{subfigure}
\caption{The generated PDFs of $Q_\textrm{sat}$ for the prior set \#1 (a) and \#2 (b).}
\label{f7}
\end{figure}

The skewness $Q_\textrm{sat}$ is the lowest order empirical parameter in SM which is almost unconstrained.
While the incompressibility modulus $K_\textrm{sat}$ is well defined, the density dependence of the incompressibility is poorly known and there are very scarce experimental analysis to determine its value.
An analysis based on charge and mass radii of the $\textrm{Sn}$ isotopes concluded that either $Q_\textrm{sat}\approx30$~MeV or $L_\textrm{sym}\approx0$~MeV~\cite{Berdichevsky1988}.
Another analysis based on the Skyrme functionals which are fitted according to the breathing modes concluded that $Q_\textrm{sat}\approx500$~MeV~\cite{Farine1997}.
A systematic analysis also suggests $Q_\textrm{sat}=300\pm400$~MeV based on a large number of theoretical models of the literature~\cite{Jerome2018p1}.

There are also other analysis based on various models from the RMF and SHF frameworks in which the EoS is constrained by using the tidal deformability of GW170817~\cite{Malik2018,Carson2019}. More precisely, the parameter $M_0$ defined as,
\begin{equation}\label{e28}
 M_0=M(n_\textrm{sat})=3n_\textrm{sat}\frac{dK(n_0)}{dn_0}\Bigg| _{n_0=n_\textrm{sat}} \, .
\end{equation}
is constrained.
The following predictions were obtained for $M_0$: $2254\leq M_0 \leq 3631$~MeV or  $1926\leq M_0 \leq 3768$~MeV depending on $L_\textrm{sym}$~\cite{Malik2018} and $1526\leq M_0 \leq 4971$~MeV~\cite{Carson2019}.

Using the relation $M_0=12K_\textrm{sat}+Q_\textrm{sat}$ (see Ref.~\cite{Alam2015}),
one can make a prediction for $Q_\textrm{sat}$ by considering adequate $K_\textrm{sat}$ value.
Considering $K_\textrm{sat}=230\pm20\enspace\textrm{MeV}$ from Ref.~\cite{Jerome2018p1}, then $-800\leq Q_\textrm{sat}\leq 1100\enspace\textrm{MeV}$ for Ref.~\cite{Malik2018} and $-1200\leq Q_\textrm{sat}\leq 2100\enspace\textrm{MeV}$ for Ref.~\cite{Carson2019}.

In Fig.~\ref{f7}, the posterior PDFs of $Q_\textrm{sat}$ are presented.
It is clear that $\chi$EFT does not constrain $Q_\textrm{sat}$.
This is because $Q_\textrm{sat}$  influences the EoS at densities well above saturation density, while the data from $\chi$EFT are relevant until $n_0=0.2$~fm$^{-3}$.
The empirical parameter $Q_\textrm{sat}$ is however better constrained by both the tidal deformability from GW170817 and the ISGMR while the predictions from prior set \#1 and \#2 are very similar.
Despite that all posteriors of tidal deformability considering TD-LVC-2018, TD-De-2018 or TD-Coughlin-2019 independently agree on the lower limit of $Q_\textrm{sat}$ ($Q_\textrm{sat}^\textrm{min}\approx -500$~MeV),
the higher boundary of $Q_\textrm{sat}$ is constrained by applying both the tidal deformability and the ISGMR constraints.
The results from joint posteriors are
$Q_\textrm{sat}=-180^{+1220}_{-175}/-160^{+935}_{-175}$~MeV ($Q_\textrm{sat}=-220^{+1130}_{-150}/-215^{+650}_{-150}$~MeV and $Q_\textrm{sat}=95^{+1365}_{-250}/200^{+1110}_{-445}$~MeV) for TD-LVC-2018 (TD-De-2018 and TD-Coughlin-2019) for the prior set \#1/\#2, respectively.

\begin{figure}
\centering
\begin{subfigure}{0.49\textwidth}
\includegraphics[width=1\textwidth]{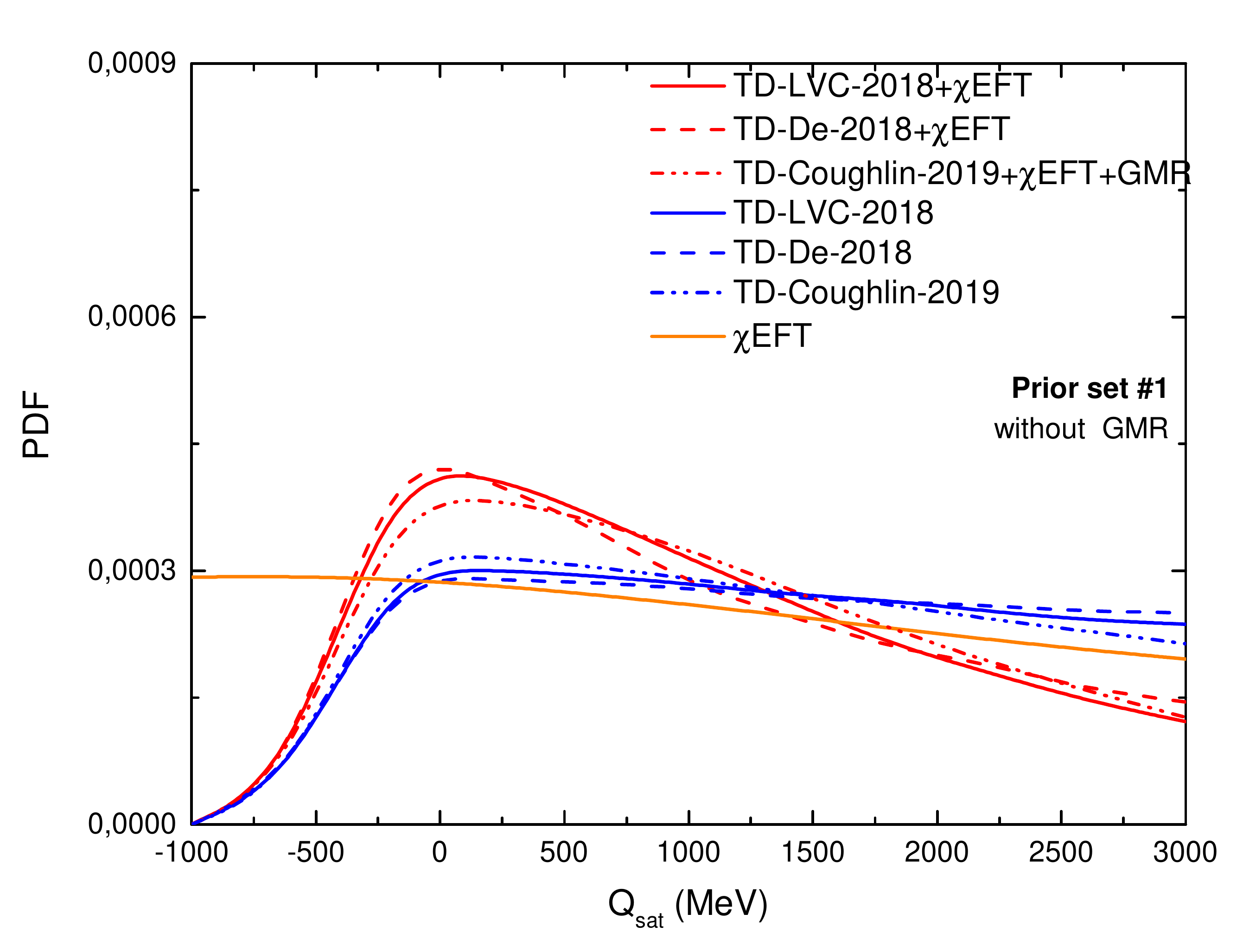}
\end{subfigure}
\caption{Same as Fig.~\ref{f7} for the prior set \#1 without ISGMR.}
\label{f7b}
\end{figure}

Furthermore, we also study the impact of switching off the ISGMR constraint for the prior set \#1 on the posterior probability in order to see its global effect on the joint posteriors, see Fig.~\ref{f7b}.
The new the joint posteriors are
$Q_\textrm{sat}=-135^{+1755}_{-250}$~MeV
($Q_\textrm{sat}=-190^{+1800}_{-200}$~MeV and $Q_\textrm{sat}=-130^{+2000}_{-250}$~MeV) for TD-LVC-2018 (TD-De-2018 and TD-Coughlin-2019).
Removing the ISGMR constraints increases the uncertainty on the joint posteriors for $Q_\textrm{sat}$ by about $500$~MeV.
This shows that $M_{c}$ is an important constraint for better defining the value of $Q_\textrm{sat}$.
Furthermore, a reduction of the uncertainty on $M_{c}$, by a systematical comparison of the meta-model predictions in finite nuclei for instance, would imply a more precise estimation for the empirical parameter $Q_\textrm{sat}$.

\subsubsection{Empirical parameter $Q_\textrm{sym}$}

\begin{figure}
\centering
\begin{subfigure}{0.5\textwidth}
\includegraphics[width=1\textwidth]{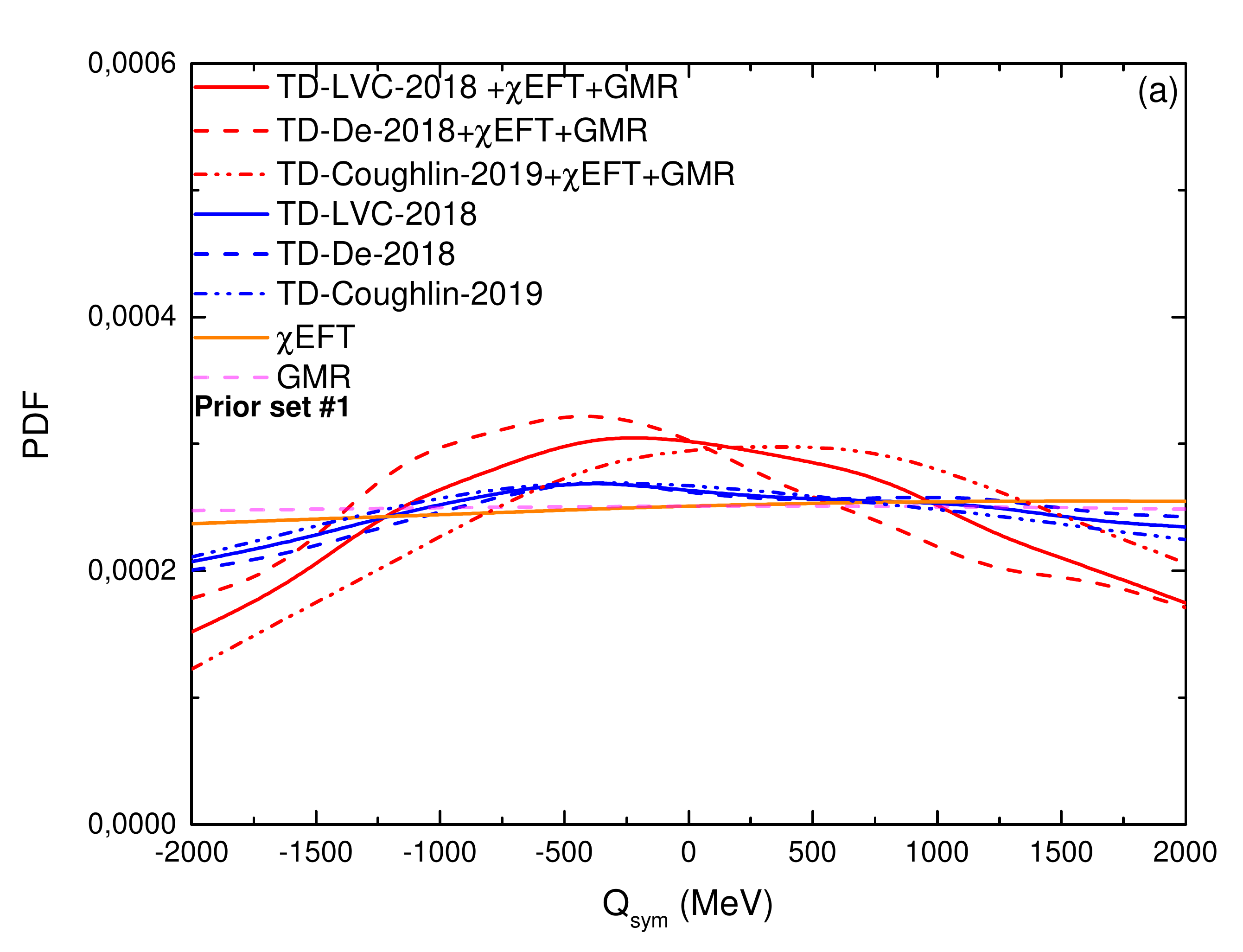}
\end{subfigure}
\begin{subfigure}{0.5\textwidth}
\includegraphics[width=1\textwidth]{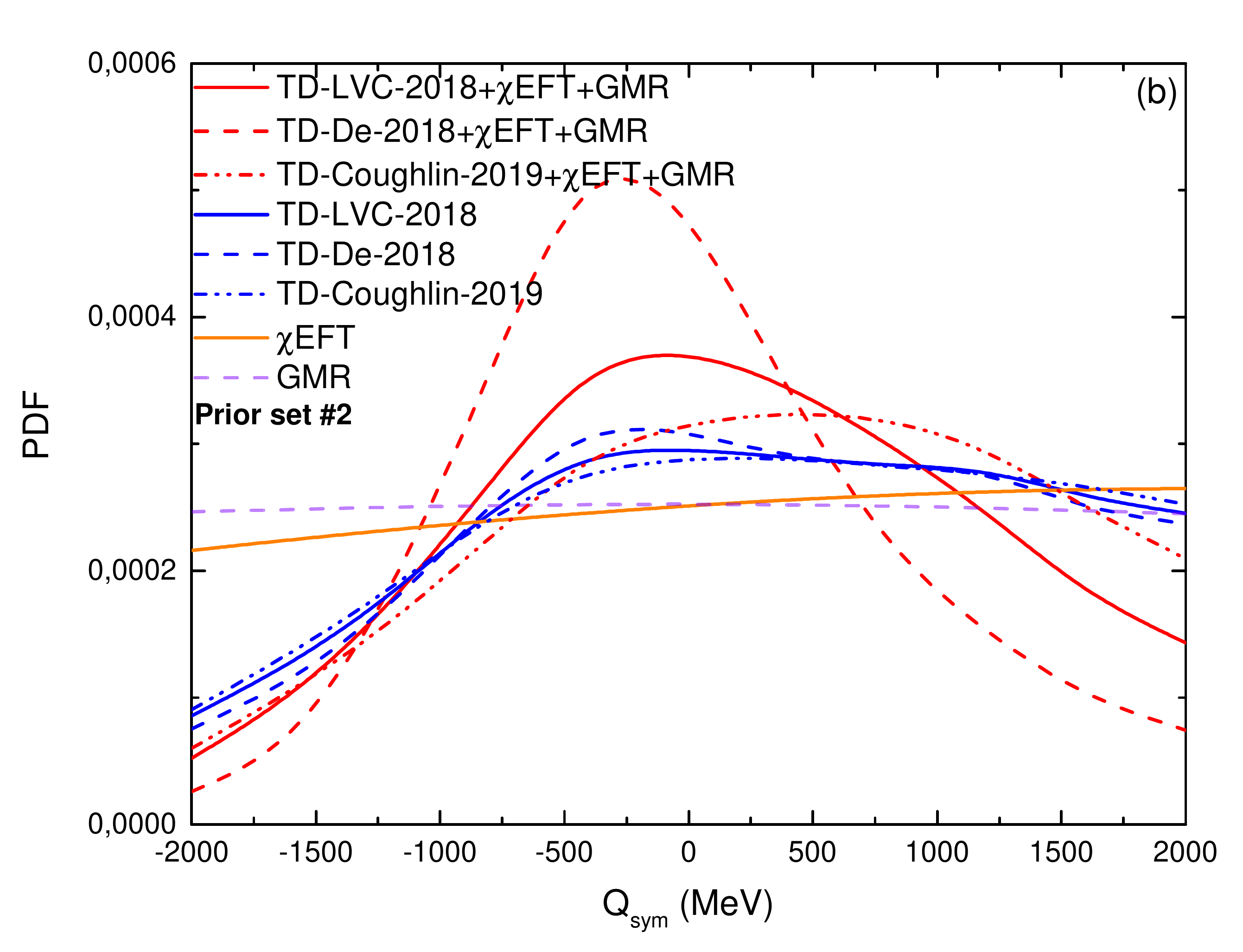}
\end{subfigure}
\caption{The generated PDFs of $Q_\textrm{sym}$ for the prior set \#1 (a) and \#2 (b).}
\label{f8}
\end{figure}

The empirical parameter $Q_\textrm{sym}$ controls the skewness of the symmetry energy at $n_\textrm{sat}$.
An analysis based on the various theoretical models (Skyrme Hartree Fock, Relativistic Hartree Fock, RMF and $\chi$EFT) suggests $Q_\textrm{sym}=0\pm400\enspace\textrm{MeV}$ but still its value runs over a large range from models to models, e.g. $-2000\leq Q_\textrm{sym}\leq 2000$~MeV~\cite{Jerome2018p1}.
Since $Q_\textrm{sym}$ contributes to the EoS at supra-saturation densities, it is quite difficult to estimate the value of this empirical parameter from low-density $\chi$EFT or from terrestrial experiments in finite nuclei like the ISGMR. It furthermore requires systems which probe asymmetric nuclear matter.
It is therefore completely unknown from nuclear physics traditional approach and one could easily understand that $\chi$EFT and ISGMR constraints are ineffective for constrain $Q_\textrm{sym}$, as shown in Fig~\ref{f8}.
The most effective constraint is provided by the tidal deformability, but it is interesting to remark that even if $\chi$EFT and ISGMR does not provide constraints taken individually, the joint posterior including tidal deformability, $\chi$EFT and ISGMR is narrower than the probability distribution considering $\tilde{\Lambda}$ alone.
The joint posteriors from TD-LVC-2018 (TD-De-2018 and TD-Coughlin-2019) favour the following values:
$Q_\textrm{sym}=-270^{+1690}_{-1125}/-170^{+1375}_{-750}$~MeV ($Q_\textrm{sym}=-675^{+1160}_{-595}/-375^{+835}_{-475}$~MeV and $Q_\textrm{sym}=220^{+1940}_{-1575}/275^{+1815}_{-1240}$~MeV)
for the prior set \#1/\#2.
It shall also be noted that there is a marked correlation between $K_\textrm{sym}$ and $Q_\textrm{sym}$: the prior set \#2 considering a tighter prior for $K_\textrm{sym}$ compared to the prior set \#1, it also predicts a narrower peak for $Q_\textrm{sym}$.
In conclusion, we point out that a more accurate PDF for $\tilde{\Lambda}$, a better constrain for $Q_\textrm{sym}$.

\subsection{Posterior probabilities for the radius $R_{1.4}$ and the pressure $P(2n_\textrm{sat})$}
\label{RP}

We now study the impact of the constraints on the posterior distribution for the NS radius at $1.4 M_\odot$, $R_{1.4}$, and the pressure at $2n_\textrm{sat}$, $P(2n_\textrm{sat})$.

\begin{figure}
\centering
\begin{subfigure}{0.5\textwidth}
\includegraphics[width=1\textwidth]{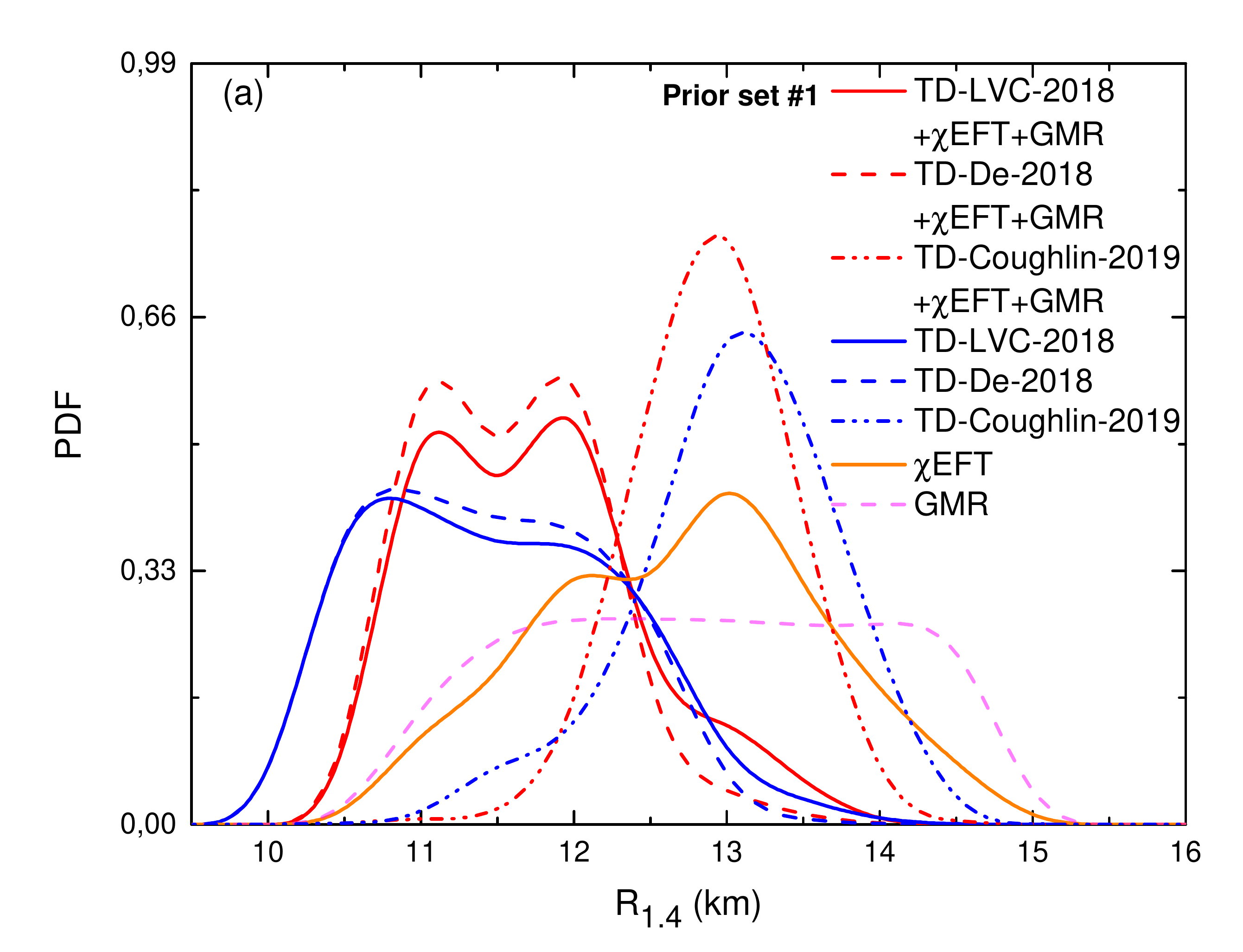}
\end{subfigure}
\begin{subfigure}{0.5\textwidth}
\includegraphics[width=1\textwidth]{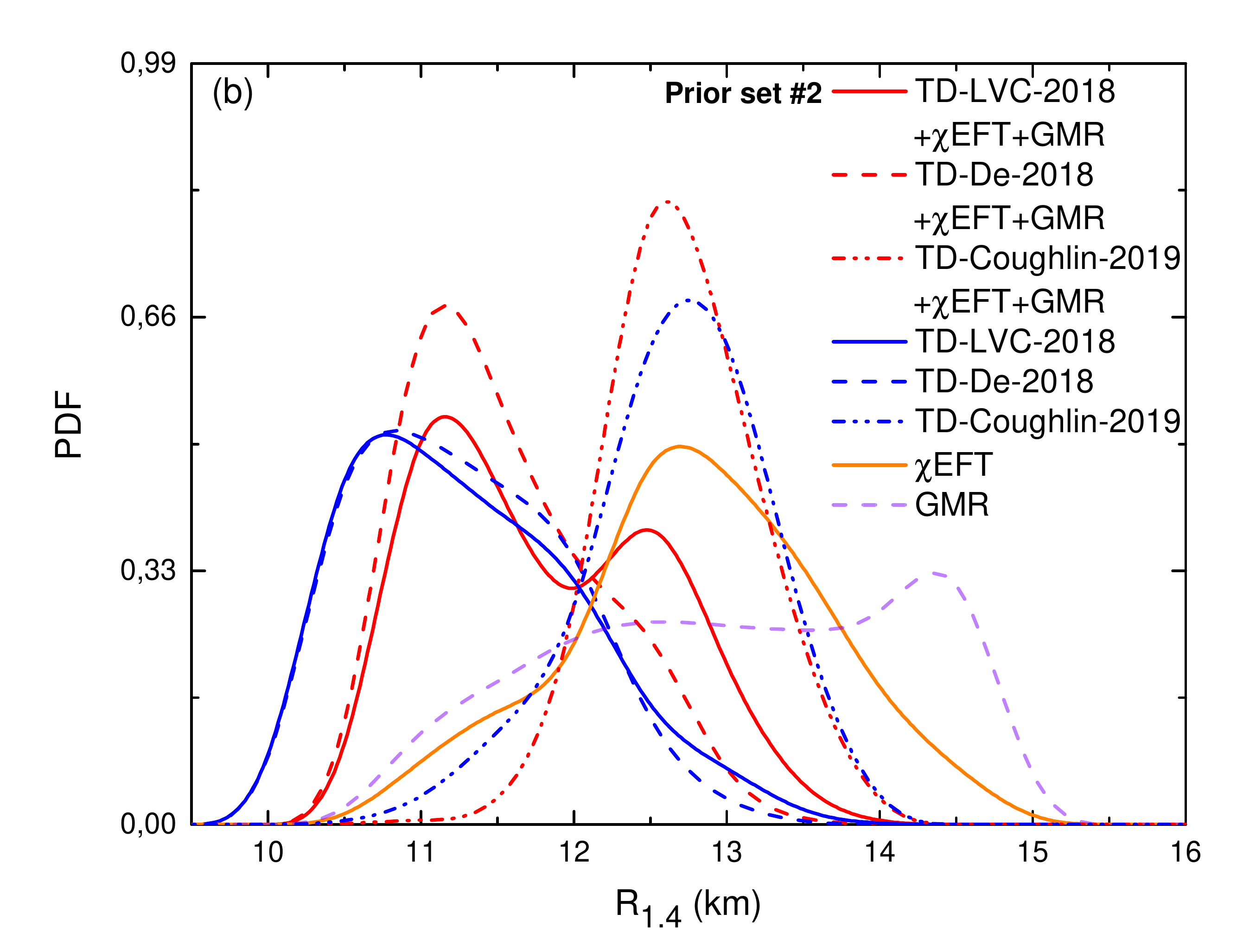}
\end{subfigure}
\caption{The generated PDFs of NS radius $R_{1.4}$ for the prior set \#1 (a) and \#2 (b).}
\label{f9}
\end{figure}

As discussed in Sec.~\ref{int}, X-ray observations of NS such as thermal emissions or X-ray bursts, advocate for
the following limits of NS radii:
$7.9\leq R_{1.4} \leq12.66$~km~\cite{Steiner2010,Steiner2014,Ozel2016,Marino2018}.
Moreover, GW analysis based on the various models concluded to $11.80$~km$\leq R_{1.4}\leq12.80$~km in Ref.~\cite{Kim2018}, $12.00$~km$\leq R_{1.4}\leq13.70$~km in Refs.~\cite{Annala2018,Most2018}, and
$11\leq R_{1.4}\leq13$~km considering $100\leq \tilde{\Lambda}\leq600$ in Ref.~\cite{Lim2018}.
While being consistent among them, these predictions are slightly different, reflecting the small model dependence in the theoretical models employed.

We show in Fig.~\ref{f9} the posteriors PDFs for the NS radius $R_{1.4}$ for the different individual constraints and for the joint one.
The predictions from TD-LVC-2018 and TD-De-2018 are $R_{1.4}=10.7^{+2.1}_{-0.3}/10.5^{+1.3}_{-0.2}$~km for the prior set \#1/\#2  at variance with the prediction from TD-Coughlin-2019 $R_{1.4}=13.1^{+0.5}_{-0.5}$~km, which are consistent with the predictions from nuclear physics ($\chi$EFT): $R_{1.4}=13.0^{+0.8}_{-1.2}/12.7^{+0.8}_{-0.6}$~km for the prior set \#1/\#2.
If the $\tilde{\Lambda}$ distribution suggested by TD-LVC-2018 and TD-De-2018 is correct, there is a difference of about 1.5~km for the most probable radii compared to the prediction from $\chi$EFT.
This difference is larger that the standard deviation for each PDF, indicating a possible source for tension, as also observed for the PDF of $L_\textrm{sym}$.
Finally, the joint probabilities shown in Fig.~\ref{f9} predict
$R_{1.4}=11.0^{+1.3}_{-0.3}/11.0^{+1.9}_{-0.3}$ or $R_{1.4}=12.0^{+0.3}_{-1.3} /11.0^{+1.9}_{-0.3}$~km
($R_{1.4}=11.0^{+1.3}_{-0.3}/11.0^{+1.7}_{-0.3}$ or
$R_{1.4}=12.0^{+0.3}_{-1.3}/11.0^{+1.7}_{-0.3}$~km and
$R_{1.4}=12.9^{+0.4}_{-0.4}/12.5^{+0.5}_{-0.3}$~km)
for TD-LVC-2018 (TD-De-2018 and TD-Coughlin-2019) for the prior set \#1/\#2.
Interestingly, the joint posteriors suggested by TD-LVC-2018 and TD-De-2018 predicts a double peak, where the first one is around $11$~km and the second one is around $12$~km for the prior set \#1.



Our prediction for $R_{1.4}$ preferred by GW170817 only (TD-LVC-2018 and TD-De-2018 but not TD-Coughlin-2019) is very similar to the one recently performed in Ref.~\cite{Capano2019}, where $R_{1.4}\approx11\pm1$~km is obtained from the analysis of the GW waveforms and the constraint from the maximum mass.
This is not entirely surprising since even if the analysis is different from ours, on the bare data in Ref.~\cite{Capano2019} and based on the post-processed analysis in terms of $\tilde{\Lambda}$ in our case, the physics issued from GW is the same.
A low value for the radius $R_{1.4}\approx 11$~km is marginal with nuclear physics (represented here by the $\chi$EFT and GMR constraints).
Our result suggest that the low peak value for $\tilde{\Lambda}\approx 200$ needs a softening of the EoS that nuclear degrees of freedom could not produce for the typical masses estimated from GW170817, which are around $1.3-1.5M_\odot$ (coinciding to central densities of about $2-3\rho_{sat}$).
This softening could be obtained by the onset of new degrees of freedom, such as pion or kaon condensation, hyperonization of matter or a first order phase transition to quark matter.
The requirement to reach about $2M_\odot$ also limits the softening, which could be obtained assuming a transition to quark matter~\cite{Montana2019}.
Such scenarios will therefore be explored in the future.

\begin{figure}
\centering
\begin{subfigure}{0.49\textwidth}
\includegraphics[width=1\textwidth]{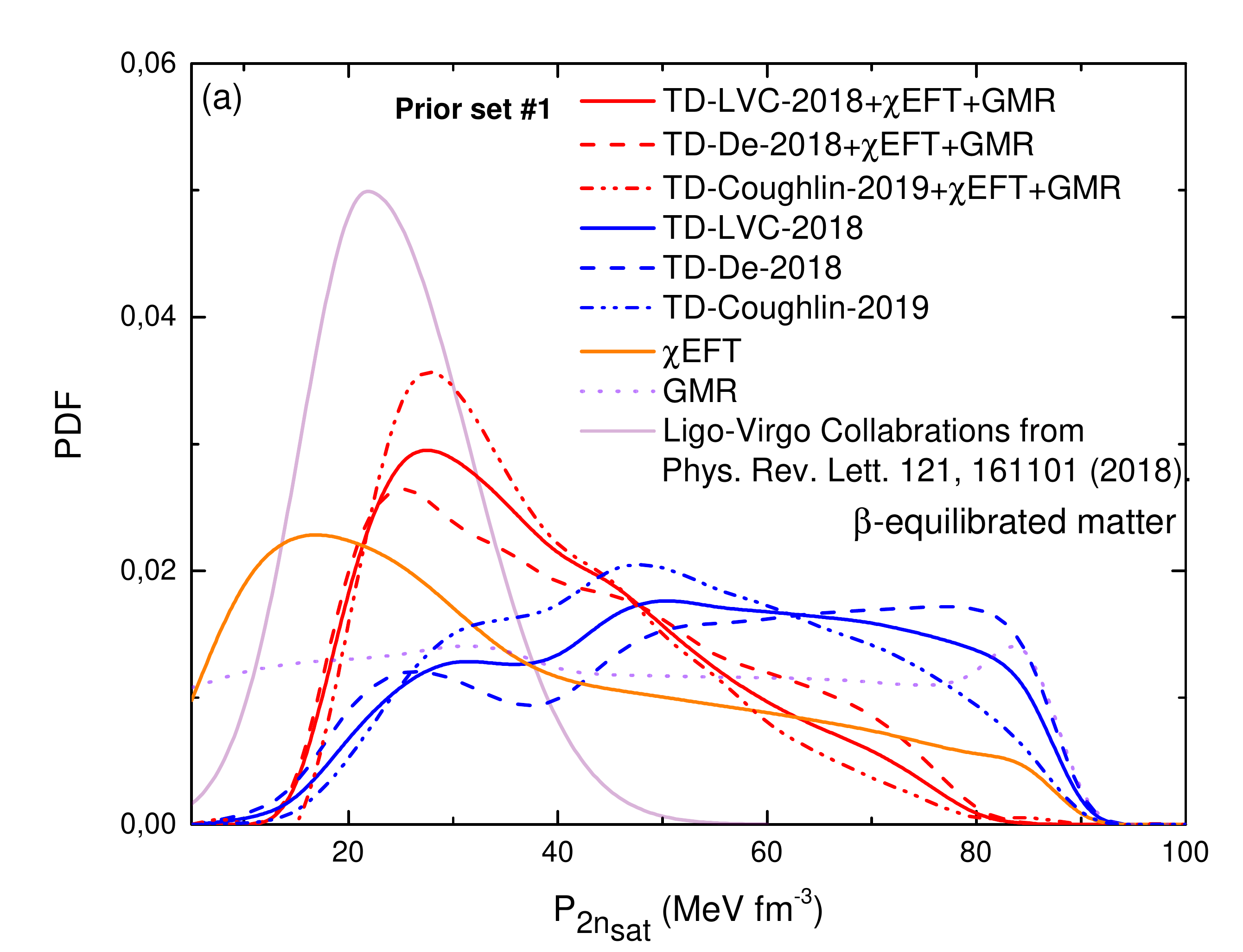}
\end{subfigure}
\begin{subfigure}{0.49\textwidth}
\includegraphics[width=1\textwidth]{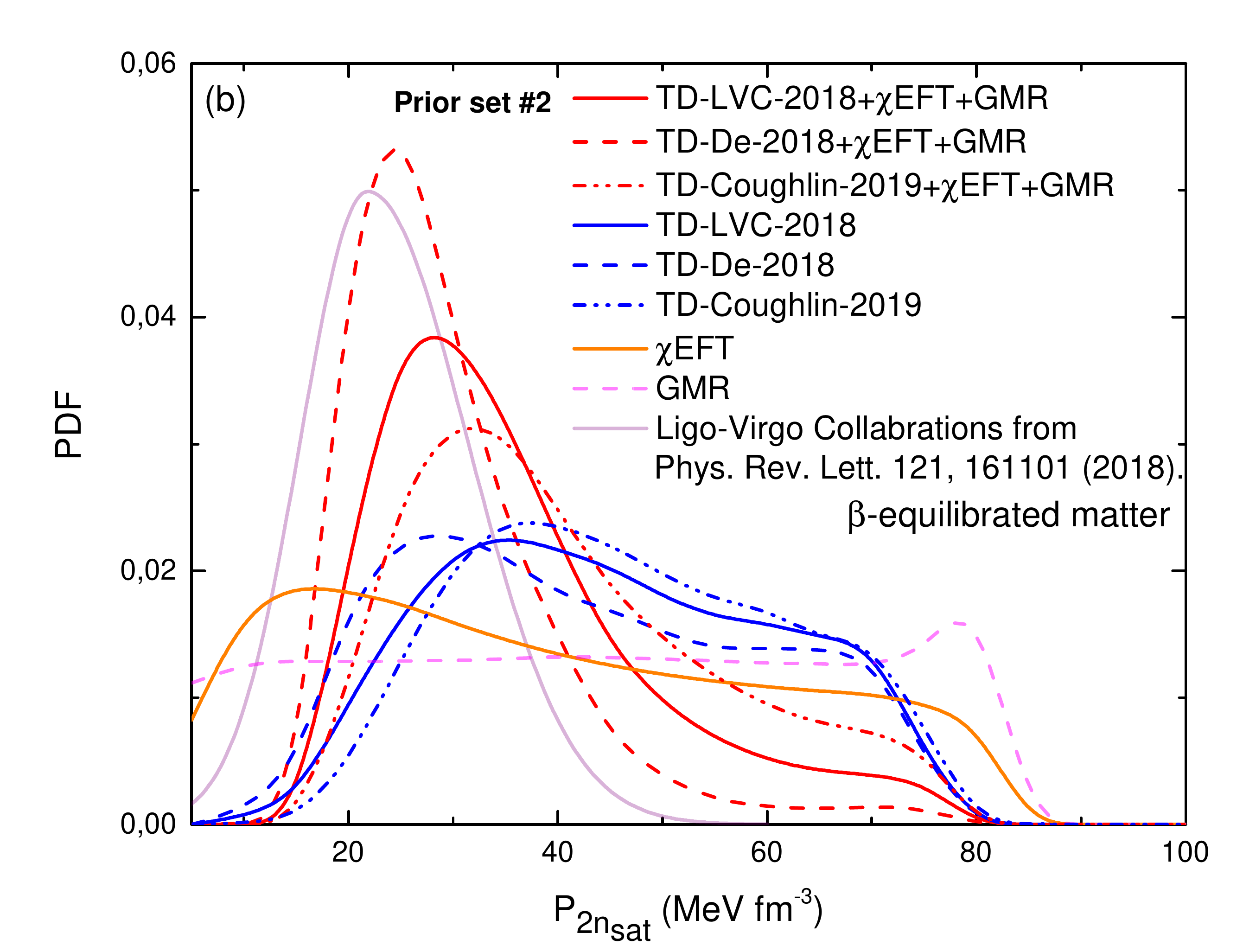}
\end{subfigure}
\caption{The generated PDFs of the pressure at $2n_\textrm{sat}$ for the prior set \#1 (a) and \#2 (b).}
\label{f10}
\end{figure}

It was recently proposed to analyze the constraint from the tidal deformability from GW170817 in terms of the pressure at $2n_\textrm{sat}$~\cite{Ligo2018}.
An analysis done by Ligo-Virgo collaborations~\cite{Ligo2018} obtained (with $90\%$ confidence interval) a pressure
$P(2n_\textrm{sat})=21.80^{+15.76}_{-10.55}$~MeV~fm$^{-3}$ where the error bars represent 90\% confidence level (corresponding to $P(2n_\textrm{sat})=21.80^{+9.58}_{-6.41}$~MeV~fm$^{-3}$ for 65\% confidence level).
Another analysis based on $\chi$EFT~\cite{Lim2018} concluded that $15\leq P(2n_\textrm{sat})\leq25$~MeV~fm$^{-3}$ considering $100\leq \tilde{\Lambda}\leq600$.

We thus further extend this approach by also imposing the nuclear physics constraints on top of the tidal deformability, in the same spirit of the previous plots (Fig.\ref{f10}).
Additionally, we have added $P(2n_\textrm{sat})$ from Ref.~\cite{Ligo2018} for comparison.
The constraints from $\chi$EFT and ISGMR generate a rather flat distribution between the boundaries with small and marginal peaks.
The tidal deformability imposes slightly stronger constraints, with $P(2n_\textrm{sat})\geq 15$~MeV for the prior set \#1 and \#2.
It is however interesting to note that here also, the joint posteriors predicts a peak narrower when including all three constraints:
$P(2n_\textrm{sat})=24.6^{+24.4}_{-5.0}/26.0^{+13.6}_{-5.0}$~MeV~fm$^{-3}$
($P(2n_\textrm{sat})=23.7^{+28.0}_{-5.0}/25.0^{+7.8}_{-5.2}$~MeV~fm$^{-3}$ and $P(2n_\textrm{sat})=25.0^{+19.9}_{-5.0}/30.0^{+18.3}_{-6.7}$~MeV~fm$^{-3}$)
for TD-LVC-2018 (TD-De-2018 and TD-Coughlin-2019) for the prior set \#1/\#2.
Although the centroid value of each tidal deformabilities are quite similar between the priors,
the prior set \#2 includes less uncertainty for TD-LVC-2018 and TD-De-2018.
Therefore, we conclude that
the limits of the pressure at $2n_\textrm{sat}$ is: $19\leq P(2n_\textrm{sat}) \leq 50$~MeV~fm$^{-3}$.
Besides,
considering the prior set \#2 which has a tighter bound for $K_\textrm{sym}$,
our prediction is in good agreement with the one proposed from Ligo-Virgo (Ref.~\cite{Ligo2018}).
The smaller dispersion is shown to come from the ISGMR, $\chi$EFT and tidal deformability considered all together.
There is however no inclusion of phase transition in the present analysis, which is expected to increase the width of the prediction~\cite{Tews2018,Tews2019}.


\subsection{Analysis of the correlations among empirical parameters}
\label{Cor}

We now present a few results on the correlations among empirical parameters originating in the different constraints investigated in this study.


\begin{figure}
\centering
\begin{subfigure}{0.49\textwidth}
\includegraphics[width=1\textwidth]{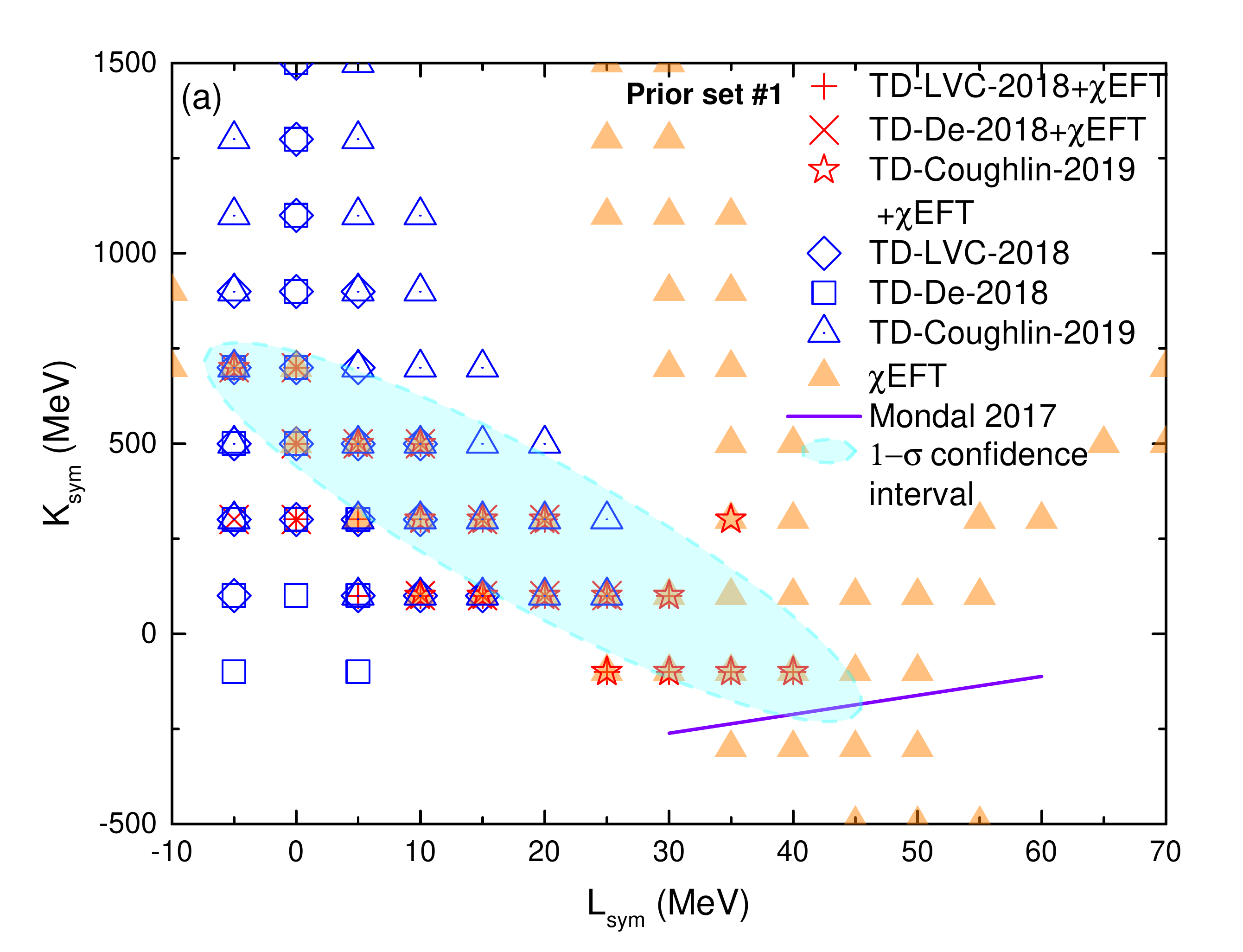}
\end{subfigure}
\begin{subfigure}{0.49\textwidth}
\includegraphics[width=1\textwidth]{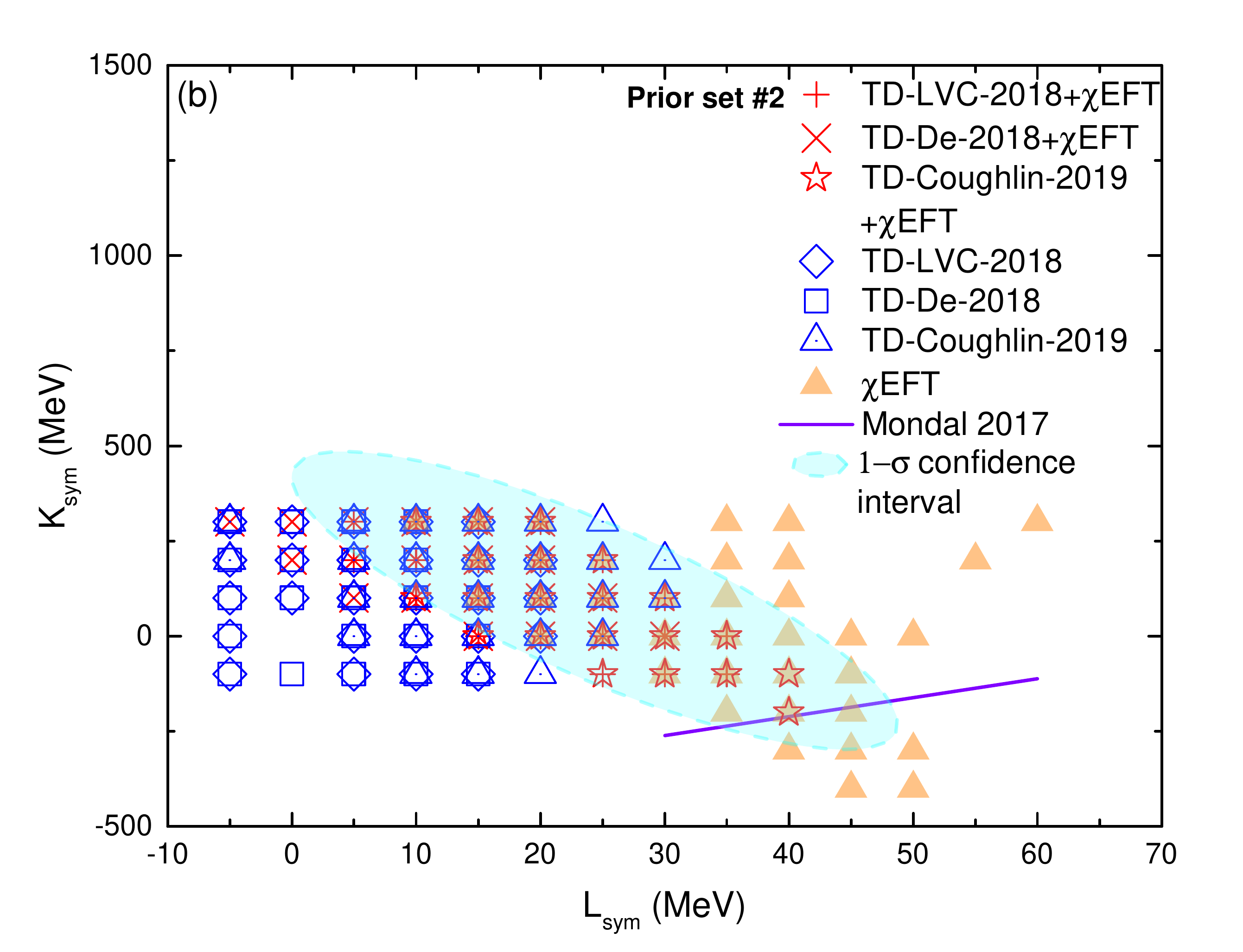}
\end{subfigure}
\caption{The values of the $L_\textrm{sym}$ and $K_\textrm{sym}$ inside of the 1-$\sigma$ probability for the prior set \#1 (a) and \#2 (b) with the fit from Ref.~\cite{Mondal2017}.}
\label{f11}
\end{figure}

\subsubsection{$L_\textrm{sym}$-$K_\textrm{sym}$ correlation}
\label{lkcor}

We first explore the correlation between $L_\textrm{sym}$ and $K_\textrm{sym}$, see Figs.~\ref{f11}, which was also explored in Refs.~\cite{Mondal2017,Chen2009,Yoshida2006,Danielewicz2009,Vidana2009,Ducoin2011,Dong2012,Santos2014}.
We remind that the influence of the prior sets on the $L_{sym}$-PDF, see Fig.~\ref{f5}, was suggesting the presence of a correlation between $L_{sym}$  and $K_{sym}$.
Here also we find a marked difference between the $L_{sym}$-$K_{sym}$ domain preferred by the GW constrain (low $L_{sym}$ values) and the one preferred by the $\chi$EFT one (high $L_{sym}$ values).
The lower bounds in $L_{sym}$ and $K_{sym}$ are imposed by the stability and $M_\textrm{max}^\textrm{obs}$ constraints, while the upper bounds are fixed by the causality one.
Note that the $L_{sym}$-$K_{sym}$ domain preferred by the TD-De-2018 $\tilde{\Lambda}$-PDF is a bit smaller than the one preferred by TD-LVC-2018.
Moreover, the prior set \#2 exploring a smaller parameter space than the prior set \#1, see Tab.~\ref{t2}, the correlation domain is smaller for set \#2 compared to \#1.
Despite this main difference, there is still a small but noticeable impact of the prior set.

Exploring a large set of RMF and Skyrme EDFs, the following relation $K_\textrm{sym}=\beta(3E_\textrm{sym}-L_\textrm{sym})+\alpha$, with
$\beta=-4.97\pm0.07$ and $\alpha= 66.80\pm2.14$~MeV, was suggested~\cite{Mondal2017}.
Fixing $E_\textrm{sym}=32$~MeV (actually $E_\textrm{sym}=32.1\pm0.3$~MeV is taken in Ref~\cite{Mondal2017}, but we keep fixed $E_\textrm{sym}=32$~MeV in our analysis,  for details see Table.~\ref{t1} and related explanations), this correlation is shown in Fig.~\ref{f11} with the caption Mondal 2017.
This correlation was shown to originate from the physical condition that the energy per particle in NM should be zero at zero density~\cite{Jerome2019}.
Using the meta-model, the validity of this correlation has been confirmed and the contribution of higher order parameter ($Q_\textrm{sym}$, $Q_\textrm{sat}$, $Z_\textrm{sym}$ and $Z_\textrm{sat}$) has also been investigated, adding about $200$~MeV uncertainty to $K_\textrm{sym}$~\cite{Jerome2019}.
There is an overlap between the Mondal 2017 correlation line and the $\chi$EFT preferred domain, as expected. However, the $\chi$EFT preferred domain is much larger since we have considered only the $n_0\ge0.12$~fm$^{-3}$ energy band in NM. The constrain at very low density is thus not included in the $\chi$EFT preferred domain.

We have also analyzed the impact of the ISGMR constraints on the $L_\textrm{sym}$-$K_\textrm{sym}$ correlation, but since this is a correlation among isovector empirical parameter, there is no impact of the ISGMR constraint.

Finally, the blue contours in Figs.~\ref{f11} represent the 1-$\sigma$ ellipses including both the GW and $\chi$EFT constraints together.
This ellipse is only weakly dependent on the prior sets \#1 and \#2.
We therefore propose a new correlation which reproduce the joint probability as,
\begin{equation}\label{s1}
K_\textrm{sym}=\alpha_{1} L_\textrm{sym}+\beta_{1} ,
\end{equation}
where $\alpha_{1}=-18.83^{+3.00}_{-2.00}$ and $\beta_{1}=616^{+140}_{-180}$~MeV.



\begin{figure}
\centering
\begin{subfigure}{0.49\textwidth}
\includegraphics[width=1\textwidth]{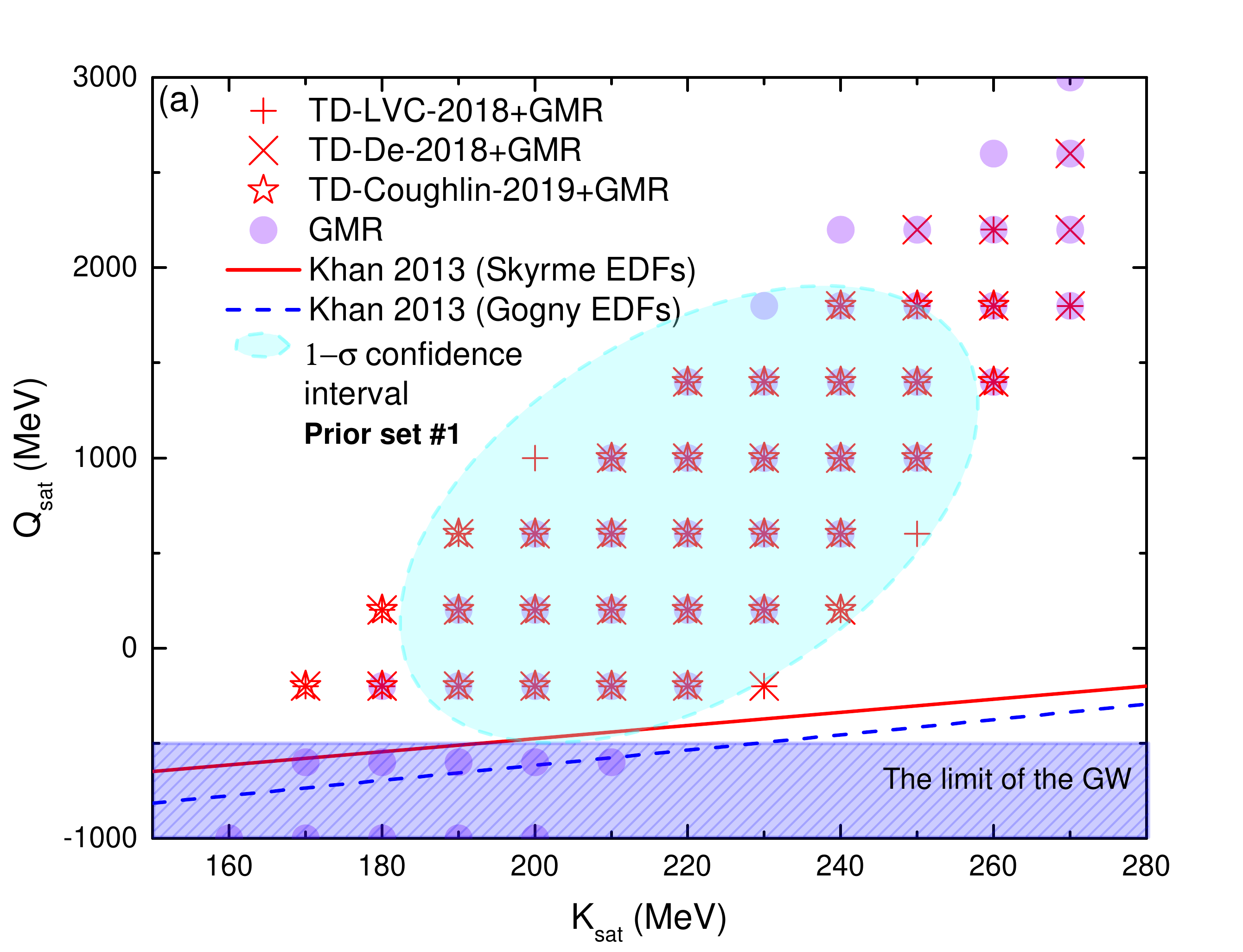}
\end{subfigure}
\begin{subfigure}{0.49\textwidth}
\includegraphics[width=1\textwidth]{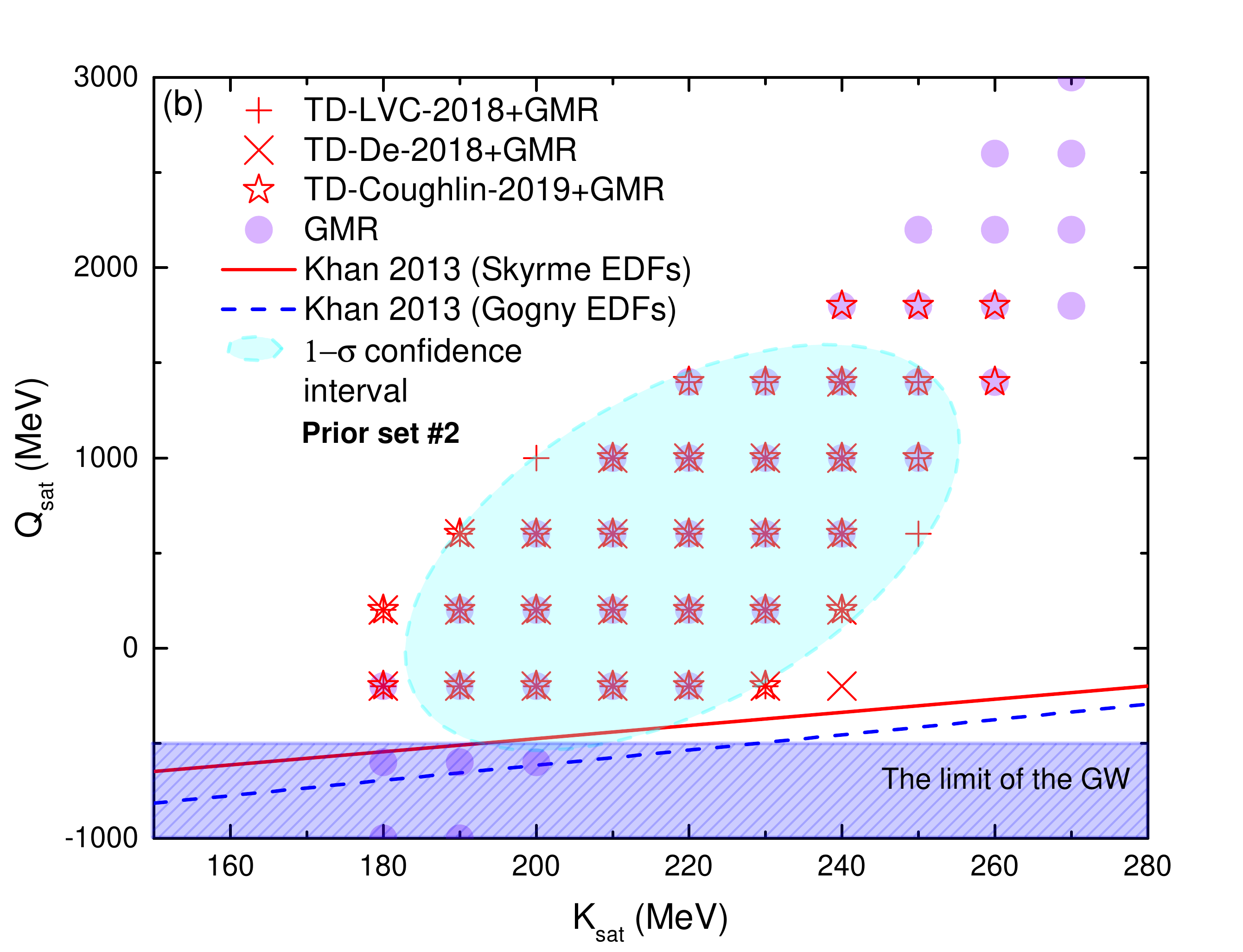}
\end{subfigure}
\caption{The values of the $K_\textrm{sat}$ and $Q_\textrm{sat}$ inside of the 1-$\sigma$ probability for the prior set \#1 (a) and \#2 (b) with a spurious correlation found for Skyrme and Gogny EDFs from Ref~\cite{Khan2013}.
Note that the $\chi$EFT constraint is included for all joint posteriors.}
\label{f12}
\end{figure}

\subsubsection{$K_\textrm{sat}$-$Q_\textrm{sat}$ correlation}

The second correlation we analyze here is the one between $K_\textrm{sat}$ and $Q_\textrm{sat}$.
The physical origin of this correlation is related to the ISGMR constraint reflected into the parameter $M_c$ defined below saturation density
at $n_c\approx 0.11$~fm$^{-3}$~\cite{Khan2012,Khan2013}.
Setting $n_0=n_c$ in the isoscalar channel ($\delta=0$) of the meta-model, one can obtain the following relation: $M_c\approx 4.6K_\textrm{sat}-0.18Q_\textrm{sat}-0.007Z_\textrm{sat}$~\cite{Jerome2019}.
Fixing $M_c=1050\pm100$~MeV, this relation induces a correlation between $K_\textrm{sat}$ and $Q_\textrm{sat}$.
However, a general analysis based on meta-model shows that this correlation is rather weak from the various EDFs and the parameter $Q_\textrm{sat}$ is yet unknown~\cite{Jerome2019}.
Since $Q_\textrm{sat}$ can be constrained by the GW data, it is worth analyzing
the correlation $K_\textrm{sat}-Q_\textrm{sat}$ under the influence of GWs.

In Figs.~\ref{f12}, the $K_\textrm{sat}$-$Q_\textrm{sat}$ correlations are shown for various constraints with a spurious correlation found for Skyrme and Gogny EDFs from Ref.~\cite{Khan2013} as the caption Khan 2013.
The source of this spurious correlation is the density dependent term from Skyrme and Gogny EDFs~\cite{Khan2013}.
First, it should be stressed that $\chi$EFT constraint is included for all joint posteriors, but its effect was found negligible in this case.
The domain allowed from the ISGMR constraint is shown with purple large dots, as previously discussed.
A lower bound $Q_\textrm{sat}\geq-500\enspace\textrm{MeV}$ is shown, originating from the GW constraint discussed in Fig.~\ref{f7}.
Finally we represent the domain allowed by the GW data with the "+" (TD-LVC-2018), "x" (TD-De-2018) and "$\star$" (TD-Coughlin-2019) symbols.
There is a nice overlap between the GW data and the ISGMR.
Furthermore, the confrontation of the GW data to the ISGMR correlation band allows us to identify a smaller domain in $K_\textrm{sat}$-$Q_\textrm{sat}$ domain, which is represented by the blue 1-$\sigma$ ellipse.
However there is a divergent result between the correlations from Skyrme and Gogny EDFs from Ref~\cite{Khan2013} and the GW since the GW favours $-500\leq Q_\textrm{sat}\leq1500\enspace\textrm{MeV}$ and
it forbids $Q_\textrm{sat}\leq-500\enspace\textrm{MeV}$.
The origin of this divergence can also be a hint for a phase transition.

From the 1-$\sigma$ confidence interval one can derive the following relation:
\begin{equation}\label{s2}
K_\textrm{sat}=\alpha_{2} Q_\textrm{sat}+\beta_{2},
\end{equation}
where $\alpha_{2}=0.035^{+0.01}_{-0.01}$ and $\beta_{2}=199^{+20}_{-30}$.
Furthermore, it seems that the GMR and GW constrain different parameter at same time.
While the GW is constraining $Q_\textrm{sat}$, the GMR impacts $K_\textrm{sat}$.
Consequently, joint posteriors predict  $170/180\leq K_\textrm{sat}\leq 250/240$~MeV and $-500/-500\leq Q_\textrm{sat}\leq 1200/1000$~MeV for the prior set \#1/\#2, respectively.
An increased resolution of both constraints shall lead to more accurate determination of $K_\textrm{sat}$ and $Q_\textrm{sat}$.

\begin{table*}[t]
\centering
\tabcolsep=0.25cm
\def\arraystretch{1.6}

\begin{tabular}{lcccccc}
\hline\hline
\multicolumn{1}{c}{\begin{tabular}[c]{@{}c@{}}Empirical Parameters \\ and \\ Neutron Star Observables\end{tabular}} & \begin{tabular}[c]{@{}c@{}}$L_\textrm{sym}$\\ (MeV)\end{tabular} & \begin{tabular}[c]{@{}c@{}}$K_\textrm{sym}$\\ (MeV)\end{tabular} & \begin{tabular}[c]{@{}c@{}}$Q_\textrm{sat}$\\ (MeV)\end{tabular} & \begin{tabular}[c]{@{}c@{}}$Q_\textrm{sym}$\\ (MeV)\end{tabular} & \begin{tabular}[c]{@{}c@{}}$R_{1.4}$\\ (km)\end{tabular} & \begin{tabular}[c]{@{}c@{}}$P(2n_\textrm{sat})$\\ (MeV fm$^{-3}$)\end{tabular} \\ \hline
TD-LVC-2018                                                                                                         & $0^{+5}_{-3}$                                                    & $375^{+\infty}_{-400}$                                           & $-101^{+\infty}_{-299}$                                          & $-521^{+\infty}_{-755}$                                          & $10.7^{+2.1}_{-0.3}$                                     & $45^{+35}_{-25}$                                           \\
TD-De-2018                                                                                                          & $0^{+2}_{-2}$                                                    & $390^{+\infty}_{-400}$                                           & $-144^{+\infty}_{-293}$                                          & $-340^{+\infty}_{-1000}$                                         & $10.7^{+2.1}_{-0.3}$                                     & $45^{+35}_{-25}$                                            \\
TD-Coughlin-2019                                                                                                    & $10^{+10}_{-10}$                                                 & $275^{+890}_{-330}$                                              & $-101^{+\infty}_{-299}$                                          & $502^{+\infty}_{-946}$                                           & $13.1^{+0.5}_{-0.5}$                                     & $45^{+35}_{-25}$                                                              \\
$\chi$EFT                                                                                                           & $35^{+7}_{-10}$                                                  & $15^{+600}_{-265}$                                                & -                                                                & -                                                                & $13^{+0.8}_{-1.2}$                                       & $12^{+23}_{-4}$                                                                \\
GMR                                                                                                                 & -                                                                & $5^{+\infty}_{-500}$                                             & $1614^{+881}_{-\infty}$                                          & $372^{+2000}_{-2000}$                                            & $12.8^{+1.5}_{-1.5}$                                     & -                                                                              \\
TD-LVC-2018+$\chi$EFT+GMR                                                                                           & $0^{+12}_{-4}$                                                   & $440^{+210}_{-210}$                                              & $-180^{+1222}_{-175}$                                            & $-271^{+1690}_{-1126}$                                           & $11^{+1.3}_{-0.2}$ or $12^{+0.3}_{-1.2}$                 & $25^{+24}_{-5}$                                                                \\
TD-De-2018+$\chi$EFT+GMR                                                                                            & $0^{+2}_{-3}$                                                    & $560^{+150}_{-150}$                                              & $-220^{+1130}_{-250}$                                            & $-677^{+1159}_{-597}$                                            & $11^{+1.2}_{-0.2}$ or $12^{+0.2}_{-1.2}$                 & $24^{+28}_{-5}$                                                                \\
TD-Coughlin-2019+$\chi$ EFT+GMR                                                                                     & $10^{+7}_{-10}$                                                  & $260^{+240}_{-240}$                                              & $93^{+1365}_{-250}$                                              & $218^{+1942}_{-1576}$                                            & $12.9^{+0.4}_{-0.4}$                                     & $25^{+20}_{-5}$                                                                \\ \hline\hline
\end{tabular}

\caption{Centroid, $\sigma_\textrm{Min}$ and $\sigma_\textrm{Max}$ values of posterior PDFs for prior set \#1.}
\label{t3}
\end{table*}

\section{Conclusions}\label{Conclusions}

In this paper, we have investigated the impact of various constraints on the EoS combining a semi-agnostic meta-model approach and Bayesian statistics.
We have analyzed the impact of the prior by comparing two different prior sets and contrasted three independent PDF for $\tilde{\Lambda}$.

Our main results are the presence of marked tensions between various analyses of the GW signal from GW170817 depending on the inclusion or absence of multi-messenger additional constraints, and marked tensions between astrophysical and nuclear physics constraints.

Let us start with the impact of the considered $\tilde{\Lambda}$-PDF.
Assuming the PDF from Refs.~\cite{De2018}~(TD-De-2018), the posteriors favors $L_\textrm{sym}=0^{+2}_{-2}$~MeV, $K_\textrm{sym}=390^{+\infty}_{-400}$~MeV, while assuming the PDF resulting from a multi-messenger analysis~(TD-Coughlin-2019)~\cite{Coughlin2019}, the posteriors favors $L_\textrm{sym}=10^{+10}_{-10}$~MeV, $K_\textrm{sym}=275^{+890}_{-330}$~MeV.
These numbers are for the prior set \# 1.
The posterior predictions based on the $\tilde{\Lambda}$-PDF from Ref.~\cite{Ligo2019}~(TD-LVC-2018) are intermediate between these two cases.
There is also a marked tension between the radius predictions $R_{1.4}$ since $R_{1.4}=10.7^{+2.1}_{-0.3}$~km for the TD-De-2018 and TD-LVC-2018 $\tilde{\Lambda}$-PDF, while it is $R_{1.4}=13.1^{+0.5}_{-0.5}$~km for the TD-Coughlin-2019 $\tilde{\Lambda}$-PDF.
Note that the radius $R_{1.4}$ predicted by the $\tilde{\Lambda}$-PDF from Ref.~\cite{De2018,Ligo2019} is consistent with another recent re-analysis of GW170817~\cite{Capano2019}.
These $\tilde{\Lambda}$-PDFs are however more consistent in their predictions for the pressure and we have found
$P(2n_\textrm{sat})=45^{+35}_{-25}$~MeV~fm$^{-3}$ for prior set \#1.

These predictions are also in marked tension with the posteriors from $\chi$EFT which predicts $L_\textrm{sym}=35^{+7}_{-10}$~MeV, $K_\textrm{sym}=15^{+600}_{-265}$~MeV, $R_{1.4}=13.0^{+0.8}_{-1.2}$~km and $P(2n_\textrm{sat})=12^{+23}_{-4}$~MeV~fm$^{-3}$ for prior set \#1.
It is interesting to remark that there is a marked tension in the values for $L_\textrm{sym}$ between all $\tilde{\Lambda}$-PDF analyses and the $\chi$EFT one.
However, for the radius $R_{1.4}$ the multi-messenger $\tilde{\Lambda}$-PDF, which is peaked at $\tilde{\Lambda}\approx 600$ is in good agreement with $\chi$EFT predictions (see Table~\ref{t3} for an overview).

Finally we have analyzed the $L_\textrm{sym}$-$K_\textrm{sym}$ and $K_\textrm{sat}$-$Q_\textrm{sat}$ correlations under the influence of GW170817, $\chi$EFT and ISGMR constraints and proposed fits for our joint probability correlations.

The tensions presented here between the posterior predictions are marked, but still consistent at 2-3$\sigma$. The reduction of the uncertainties in our predictions requires a reduction of the observational or experimental uncertainties.
So increasing the accuracy on the determination of  tidal deformability from GW, as well as $M_c$ from the ISGMR, will lead to a better determination of $K_\textrm{sat}$ and $Q_\textrm{sat}$ and NS properties.
Increasing the number of BNS GW signals is also a way to refine our present analysis and conclude on the strength of the tension between multi-physics constraints.

\section{Acknowledgement}

This work is supported by the Yildiz Technical University under project number FBI-2018-3325,
Technological Research Council of Turkey (T\"{U}B\.{I}TAK) B\.{I}DEB 2214-A Doctoral Research program and it
is partially supported by the Scientific and Technological Research Council of Turkey (T\"{U}B\.{I}TAK)
under project number MFAG-118F098.
The authors are grateful to the LABEX Lyon Institute of Origins (ANR-10-LABX-0066) of the Universit\'e de Lyon for its financial support within the program "Investissements d'Avenir" (ANR-11-IDEX-0007) of the French government, operated by the National Research Agency (ANR). EK and JM were partially supported by the IN2P3 Master Project MAC. The authors also thank the "NewCompStar" COST Action MP1304 and PHAROS COST Action MP16214.


%

\end{document}